%% file: HiggsMassQuarkr.tex
\documentclass[a4paper,12pt]{article}
\pdfoutput=1 % if your are submitting a pdflatex (i.e. if you have
\usepackage{jheppub} % for details on the use of the package, please
\usepackage[T1]{fontenc} % if needed

\usepackage[latin1]{inputenc}
\usepackage[english]{babel}

\usepackage{amssymb,amsthm,cancel,hyperref,graphicx,xcolor}
\usepackage{picinpar,graphicx,xy}
\usepackage[subnum]{cases}
\usepackage{booktabs}
\usepackage{mathrsfs}
\usepackage{amsfonts}
\usepackage{latexsym}
\usepackage{booktabs,graphicx,hyperref,epsfig,multirow}
\usepackage{bm}
\usepackage{comment}
\allowdisplaybreaks
\def\twiddles#1{\mathrel{\mathop{\sim}\limits_
                        {\scriptscriptstyle {#1\rightarrow 0 }}}}
\def\twiddlesinf#1{\mathrel{\mathop{\sim}\limits_
                        {\scriptscriptstyle {#1\rightarrow \infty }}}}
\def\bea{\begin{align}}
\def\eea{\end{align}}
\def\beq{\begin{equation}}
\def\eeq{\end{equation}}
\def\ba{\begin{eqnarray}}
\def\ea{\end{eqnarray}}
\def\be{\begin{equation}}
\def\ee{\end{equation}}
\definecolor{darkgreen}{HTML}{008000}
\newcommand{\sss}{\scriptscriptstyle\rm}
\newcommand{\muf}{\mu_{\rm\sss F}}
\newcommand{\mur}{\mu_{\rm\sss R}}

\newcommand{\Li}{\mathrm{Li}}

\newcommand{\abs}[1]{\left|\,#1\,\right|}

\newcommand{\Ord}{\mathcal{O}}
\newcommand{\gsim}{\gtrsim}
\newcommand{\lsim}{\lesssim}
\newcommand{\as}{\alpha_s}

\newcommand{\barxi}{\bar{\xi}}

\def\({\left(}
\def\){\right)}
\def\[{\left[}
\def\]{\right]}
\def    \hepph  #1 {{\tt hep-ph/#1}}
\def    \hepex  #1 {{\tt hep-ex/#1}}
\long\def\symbolfootnote[#1]#2{\begingroup%
\def\thefootnote{\fnsymbol{footnote}}\footnote[#1]{#2}\endgroup}

\numberwithin{equation}{section}
\def\lapprox{\lower .7ex\hbox{$\;\stackrel{\textstyle <}{\sim}\;$}}
\def\gapprox{\lower .7ex\hbox{$\;\stackrel{\textstyle >}{\sim}\;$}}

\renewcommand{\(}{\left(}
\renewcommand{\)}{\right)}

\newcommand{\Ca}{C_{\rm\sss A}}

\newcommand{\Gf}{G_{\sss F}}

\newcommand{\yt}{y_{t}}
\newcommand{\yb}{y_{b}}

\newcommand{\pt}{p_{\rm\sss T}}

\graphicspath{{Images/}}
\definecolor{darkblue}{rgb}{0,0,0.7}

\begin{document}
\begin{flushleft}
%%%%%%%%%%%%%%%%%%%%%%%%%%%%%%%%%%%%%%%%%%%%%%%%
\begin{figure}[h]
\includegraphics[width=.2\textwidth]{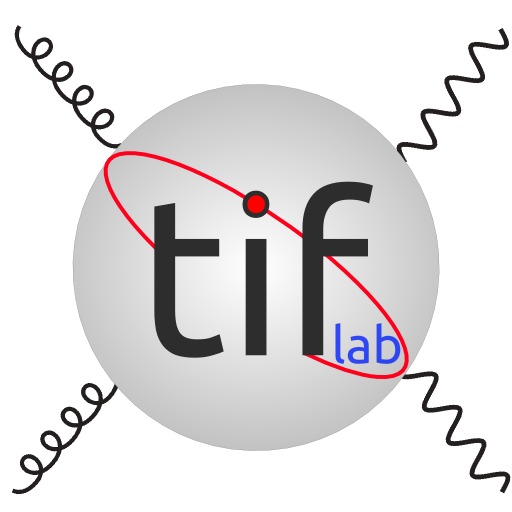}
\end{figure}
%%%%%%%%%%%%%%%%%%%%%%%%%%%%%%%%%%%%%%%%%%%%%%%%%
\end{flushleft}
\vspace{-5.0cm}
\begin{flushright}
CERN-TH-2016-139\\
TIF-UNIMI-2016-4\\
MIT-CTP/4799\\
\end{flushright}

\vspace{2.0cm}

\begin{center}
{\Large \bf 
The Higgs transverse momentum spectrum with finite quark masses beyond leading order}
\end{center}

%\vspace{1.3cm}

\begin{center}
Fabrizio Caola$^1$, Stefano Forte$^2$, Simone Marzani$^3$,\\ Claudio
Muselli$^2$ and Gherardo Vita$^4$. \\
\vspace{.3cm}
{\it
{}$^1$ Theory Division, CERN, CH-1211 Gen\`eve 23, Switzerland\\
{}$^2$Tif Lab, Dipartimento di Fisica, Universit\`a di Milano and\\
INFN, Sezione di Milano, Via Celoria 16, I-20133 Milano, Italy\\
{}$^3$Department of Physics, University at Buffalo,\\ The State University of New York
Buffalo, NY 14260-1500, USA\\
{}$^4$Center for Theoretical Physics, Massachusetts Institute
of Technology,\\ Cambridge, MA 02139, USA
}
\end{center}

\begin{center}
{\bf \large Abstract}
\end{center}
We apply the leading-log high-energy resummation technique recently
derived by some of us to the transverse momentum distribution for
production of a Higgs boson in gluon fusion. We use our results to
obtain information on mass-dependent corrections to this observable,
which is only known at leading order when exact mass dependence is
included.  In the low $\pt$ region we discuss the all-order
exponentiation of collinear bottom mass logarithms. In the high $\pt$
region we show that the infinite top mass approximation loses accuracy
as a power of $\pt$, while the accuracy of the high-energy
approximation is approximately constant as $\pt$ grows. We argue
that a good approximation to the NLO result for $p_T\gsim200$~GeV can
be obtained by combining the full LO result with a $K$-factor computed
using the high-energy approximation.
\clearpage
\tableofcontents
\clearpage

\section{Introduction}
The discovery of the Higgs boson at the CERN Large Hadron Collider
(LHC)~\cite{Aad:2012tfa,Chatrchyan:2012xdj} set an important milestone
for our understanding of fundamental interactions. So far, the
properties of the new particle seem consistent with Standard Model
predictions, which suggests a simple electroweak symmetry breaking
sector~\cite{Aad:2015gba,Khachatryan:2014jba}. A major goal of the LHC
Run II is establish
whether the new particle is indeed the Higgs Boson of the Standard
Model or there are some deviations pointing towards new physics. In
order to
reach this goal, very precise theoretical
predictions  for signal and
background processes are mandatory.

The dependence of the cross-section on heavy quark masses is
an interesting probe of the properties of the Higgs boson
both within and beyond the Standard Model, since it gives access
to
 the structure of the $ggH$ coupling~\cite{Arnesen:2008fb,Dolan:2012rv,Grojean:2013nya,Englert:2013vua,Harlander:2013oja,Wiesemann:2015fxy}. 
A particularly interesting observable in this respect is the
transverse momentum distribution, since it allows a study of the
$ggH$ coupling at different energy scales and can then provide
valuable information on its structure. 

Gluon fusion is the main Higgs production mechanism at the LHC. In this channel the total production cross-section was
recently computed to N$^3$LO
accuracy~\cite{Anastasiou:2015ema,Anastasiou:2016cez}. 
Recently, fully differential results for Higgs production in
association with one hard jet have become
available~\cite{Boughezal:2015dra,Boughezal:2015aha,Chen:2014gva}.
However, all these results have been obtained in the approximation in which
heavy quark masses are assumed to be very large,
and the coupling of the Higgs boson to gluons is
then described using an effective theory.

At the inclusive level, this is just as well since the dependence on
the heavy quark mass is very weak and under good
theoretical control at present collider energies~\cite{Harlander:2009my}.  On the other hand, large
effects are expected in the transverse momentum distribution.  Indeed,
theoretical predictions for this observable in the full theory, which
are only known at the lowest nontrivial order~\cite{Baur:1989cm}, show
large deviations from the effective theory as soon as $\pt$ is
comparable to the top quark mass.  
The fact that only the leading order is known 
is particularly problematic since we know from the inclusive case 
that radiative corrections are very large.

In Ref.~\cite{Marzani:2008az} some of us have shown that using
high-energy resummation methods it is possible to glean partial
information on the heavy quark mass dependence at higher order, at the
level of the inclusive cross-section. These results were subsequently
used to construct an optimized approximation to  the
NNLO~\cite{Harlander:2009mq,Harlander:2009my,Pak:2009dg} and N$^3$LO~\cite{Ball:2013bra,Bonvini:2014jma,Bonvini:2016frm,Harlander:2016hcx}
inclusive cross-section with full top mass dependence.
The goal of this paper is to apply similar ideas to transverse
momentum distributions; this is possible thanks to the recent
derivation~\cite{Forte:2015gve} of 
high-energy resummation for  transverse momentum
distributions. 

High-energy resummation is available only at the
leading logarithmic level: it provides us with information on the
contribution to all orders in $\alpha_s$ which carries the highest
logarithmic power of $\ln\frac{s}{m_h^2}$. Still, this  provides relevant 
insight on the heavy quark mass dependence. Indeed,
 in the opposite kinematic limit, namely the threshold limit
in which $\frac{m_h^2}{s}\to 1$, all the dependence on the heavy
quark mass can be absorbed in a factorized Wilson coefficient which
depends only on the strong coupling and the ratio of the heavy quark
to the Higgs mass, up to terms suppressed by powers of
$1-\frac{m_h^2}{s}$. On the contrary, in the high-energy limit the behaviour of
the total cross-section in the effective and full theory are qualitatively
different, as the former is double-logarithmic~\cite{Hautmann:2002tu} and the latter
single-logarithm~\cite{Marzani:2008az} (i.e. they are respectively a series in 
 $\alpha_s\ln^2\frac{s}{m_h^2}$ and $\alpha_s\ln \frac{s}{m_h^2}$). 

In Ref.~\cite{Forte:2015gve}, where a general resummation of
transverse momentum distributions was derived, a first application to
Higgs production in gluon fusion in the effective field theory limit
was presented. Here, we will apply the same general formalism to the
same observable, but now retaining full heavy quark mass
dependence. Besides studying the top mass dependence in the boosted
Higgs region, our results provide some insight on  bottom logs when
both top and bottom mass dependence is retained.  Indeed,
 in the region $m_b^2 \lsim \pt^2$ bottom mass
effects may become relevant. Of particular interest is the region
$m_b^2 < \pt^2 < m_h^2$, in which logarithmically enhanced, though
mass-suppressed terms
appear~\cite{Banfi:2013eda}. We
will be able to study these logs to all orders, albeit in the high-energy limit.

The paper is organized as follow: in Sect.~\ref{sec:resummation} we
present the resummation of the transverse momentum distribution for
Higgs production in gluon fusion with complete quark mass dependence.
In Sect.~\ref{sec:parton} we discuss the partonic resummed
cross-section. We check that its leading order truncation agrees with
the high-energy limit of the exact result,
and that in the pointlike limit it reproduces the resummed result of
Ref.~\cite{Forte:2015gve}.  We study the first few orders of its
perturbative expansion, and specifically  we 
study the high-$\pt$ region and
compare the high-energy result expanded through NLO in the effective
and full theory. We use these result as a way to qualitatively estimate
mass corrections beyond leading order: we show that for
high enough transverse momenta the high-energy approximation provides
a reasonable estimate of higher-order corrections while the effective
field theory fails completely.
We also  address to all orders the structure of
the logarithmic dependence on the bottom mass. In
Sect.~\ref{sec:hadro} we discuss phenomenological implications: we repeat the comparison of various
approximations of  Sect.~\ref{sec:parton} but now at the level of hadronic
cross-sections and $K$-factors. We conclude that currently the best
approximation in the high $p_T\gsim200$~GeV region is obtained by
combining the exact LO result with a $K$-factor determined in the
high-energy approximation. We also compare our results to previous
estimates of finite mass effects based on matching to
parton showers~\cite{Hamilton:2015nsa}.  More accurate approximations could be
obtained by combining multiple resummations, as we discuss in
Sect.~\ref{sec:conclusion} where conclusions are drawn and future
developments are discussed.

\section{Resummation}
\label{sec:resummation}

Leading-log high-energy resummation has been known for inclusive
cross-sections~\cite{Catani:1990xk,Catani:1990eg} and rapidity
distributions~\cite{Caola:2010kv} since a long time. More recently,
a framework for the resummation of transverse-momentum spectra was
developed by some of us~\cite{Forte:2015gve}. In this section,
after a brief summary of notation and conventions,
we apply it to the resummation of the Higgs transverse momentum distribution
in gluon fusion  with finite top and bottom masses, and then study its
perturbative expansion, which will allow us to obtain the truncation of
the resummed result to any finite order.

\subsection{Kinematics and definitions}
\label{subsec:kin}

In standard collinear factorization, the hadron-level transverse momentum
distribution can be written as
\begin{multline}
\label{eq:fact1}
\frac{d\sigma}{d\xi_p}\(\tau,\xi_p,\{y_i\}\)=\sum_{ij}\int_{\tau\(\sqrt{1+\xi_p}+\sqrt{\xi_p}\)^2}^1 dx_1 
\int_{\frac{\tau\(\sqrt{1+\xi_p}+\sqrt{\xi_p}\)^2}{x_1}}^1 dx_2 \\ \times
\frac{d\bar{\sigma}_{ij}}{d\xi_p}\(\frac{\tau}{x_1
  x_2},\xi_p,\{y_i\},\as(\mur^2),\mur^2,\muf^2\)
 f_i\(x_1,\muf^2\)f_j\(x_2,\muf^2\), 
\end{multline}
where $f_i(x_i)$ are parton distributions and we 
parametrized the kinematics in terms of the following dimensionless 
ratios
\beq
\label{eq:xis}  \tau=\frac{m_h^2}{s};\> 
\xi_p=\frac{\pt^2}{m_h^2}; \> 
y_i=\frac{m_i^2}{m_h^2} \eeq where $m_h$, $m_i$ are respectively the
Higgs and the various heavy quark masses, $\pt$ is the transverse momentum of the outgoing Higgs
boson and $s$ is the (hadronic) center-of-mass energy.

Equation~(\ref{eq:fact1}) can be cast in the form
of a standard convolution by an appropriate choice of hard scale. To
see this, we define 
\beq
\label{eq:tauprime}
\tau'(\tau,\xi_p)=\tau\(\sqrt{1+\xi_p}+\sqrt{\xi_p}\)^2= \frac{Q^2}{s},
\eeq
thus identifying the threshold energy
\beq
\label{eq:physcale}
\sqrt{Q^2}=\sqrt{m_H^2+\pt^2}+\sqrt{\pt^2}
\eeq
with the physical scale of the process. Note that when $\pt\ll m_H$,
$\tau'\approx\tau$, while when $\pt\gg m_H$, $\tau'\approx \frac{4\pt^2}{s}$.
If we now introduce the partonic equivalent of Eq.~(\ref{eq:tauprime})
\beq\label{eq:xip}
x'=\frac{Q^2}{\hat s},
\eeq
we can rewrite the hadronic cross-section as 
\beq
\label{eq:fact2}
\frac{d\sigma}{d\xi_p}\(\tau,\xi_p,\{y_i\}\)=\tau' \sum_{ij}
\int_{\tau'}^1
\frac{dx'}{x'}\mathcal{L}_{ij}\(\frac{\tau'}{x'},\muf^2\)\left[\frac{1}{x'}
\frac{d\hat{\sigma}_{ij}}{d\xi_p}\(x',\xi_p,\{y_i\},\as(\mur^2),\mur^2,\muf^2\)\right],
\eeq
where 
the parton luminosity is defined in the usual way as
\beq
\mathcal{L}_{ij}\(x,\muf^2\)=\int_{x}^1 \frac{dy}{y} f_i\(y,\muf^2\)f_j\(\frac{x}{y},\muf^2\),
\eeq
and 
\beq
\label{eq:Cpthadr}
\frac{d\hat{\sigma}_{ij}}{d\xi_p}\(x',\xi_p,\{y_i\},\as,\mur^2,\muf^2\)=\frac{d\bar{\sigma}_{ij}}{d\xi_p}\(\frac{x'}{\(\sqrt{1+\xi_p}+\sqrt{\xi_p}\)^2},\xi_p,\{y_i\},\as,\mur^2,\muf^2\).
\eeq
That $Q^2$ is a natural choice for the process is
demonstrated by the fact that Eq.~(\ref{eq:fact2}) takes the form of a
convolution, and thus in particular it factorizes upon Mellin
transformation. In the following, we fix $\mur^2=\muf^2=Q^2$ and drop
for simplicity the dependencies on these scales. 
The full scale dependence can be restored at any stage
using renormalization group arguments.

High-energy resummation is usually performed in Mellin ($N$) space. 
            For the sake of the determination of the leading-logarithmic
(LL$x$) result, it is immaterial whether the scale is chosen as $Q^2$
Eq.~(\ref{eq:physcale}) (so the Mellin $N$ variable is conjugate to
$\tau'$ Eq.~(\ref{eq:tauprime})) or $m_H^2$ (so
Mellin $N$ is conjugate to $\tau$ Eq.~(\ref{eq:xis})), because the
choice of scale is a subleading $\ln x$ effect.
The LL$x$ expression of the partonic cross-section can be expressed in terms of the Mellin
transform\footnote{Note that with a slight abuse of notation we use
the same notation for a function and its Mellin transform.}
\beq
\label{eq:melpt}
\frac{d{\hat\sigma}_{ij}}{d \xi_p}\(N,\xi_p,\{y_i\},\as\)=\int_0^1 dx\,
x^{N-1} \frac{d\hat{\sigma}_{ij}}{d\xi_p}\(x,\xi_p,\{y_i\},\as\) \eeq
through an impact factor $h_{\pt}$: 
\beq
\label{eq:sigmaresxp}
\frac{d\hat\sigma_{ij}}{d\xi_p}\(N,\xi_p,\{y_i\},\as\)=
h_{ij,\pt}\(0,\gamma\(\frac{\as}{N}\),\gamma\(\frac{\as}{N}\),\xi_p,\{y_i\}\),
\eeq where $\gamma\(\frac{\as}{N}\)$ is the BFKL LL$x$ resummed
anomalous
dimension~\cite{Lipatov:1976zz,Fadin:1975cb,Kuraev:1976ge,Kuraev:1977fs,Balitsky:1978ic,Altarelli:1999vw}.
The impact factor for the $gg$ channel is defined as
\begin{multline}
\label{eq:hstart}
h_{gg,\pt}\(N,M_1,M_2,\xi_p,\{y_i\}\)=h_{\pt}\(N,M_1,M_2,\xi_p,\{y_i\}\)=M_1
M_2 R\(M_1\) R\(M_2\)\times\\ \times\int_0^\infty d\xi\,\xi^{M_1-1}\int_0^\infty
d\bar{\xi}\,\bar{\xi}^{M_2-1}\,C_{\pt}\(N,\xi,\bar{\xi},\xi_p,\{y_i\}\).
\end{multline}
Here the process-dependent coefficient function $C_{\pt}$ describes
the interaction of two hard off-shell gluons with the Higgs boson (its
computation will be described in the next Subsection), the Mellin
transforms in $\xi$ and $\bar \xi$ resum multiple high-energy gluon
emission and $R(M)$ is a function which fixes the factorization
scheme; the reader is referred to Ref.~\cite{Forte:2015gve} for full
derivations and details.

Due to the eikonal nature of high-energy gluon evolution,
results for all other partonic channels can be trivially obtained
from Eq.~(\ref{eq:hstart}):
\begin{align}
\label{eq:hqq}
h_{qg,\pt}(N,M_1,M_2,\xi_p,\{y_i\}) =&
\frac{C_F}{C_A}\bigg[ h_{\pt} (N,M_1,M_2,\xi_p,\{y_i\}) -
h_{\pt}(N,0,M_2,\xi_p,\{y_i\})\bigg],\\ 
h_{qq',\pt}(N,M_1,M_2,\xi_p,\{y_i\})=&
\left(\frac{C_F}{C_A}\right)^2
\bigg[ h_{\pt}(N,M_1,M_2,\xi_p,\{y_i\}) + \nonumber \\&-
h_{\pt}(N,0,M_2,\xi_p,\{y_i\})-
h_{\pt}(N,M_1,0,\xi_p,\{y_i\})],
\end{align}
where $q,q'$ can be any quark or anti-quark. The subtraction
terms in Eq.~(\ref{eq:hqq}) ensure that at least one emission
from the quark line is present, see Ref.~\cite{Forte:2015gve} for
details; note that  subtraction of $h_{\pt}(N,0,0,\xi_p,\{y_i\})$ is
not necessary because this contribution vanishes for $\pt \not=0$.

\subsection{The impact factor}
\label{subsec:evalimpact}
The computation of the coefficient function $C_{\pt}$ which enters
Eq.~(\ref{eq:hstart}) follows the procedure outlined in
Refs.~\cite{Marzani:2008az,Forte:2015gve}:  
$C_{\pt}$ is closely related to 
the transverse momentum distribution for the process
\beq
g^*\(k_1\)+g^*\(k_2\) \to H\(p\).
\eeq
Specifically, the off-shell gluon momenta can be parametrized in terms
of longitudinal 
and transverse components as
\begin{align}\label{eq:zzbardef}
&k_1 = z p_1 + k_{t,1}\nonumber\\
&k_2 = \bar z p_2 + k_{t,2}
\end{align}
with
\begin{align}\label{eq:vardefs}
&p_i^2 = 0,\quad p_i \cdot k_{t,j} = 0,\quad i,j=1,2 \nonumber\\
&k_1^2 = k_{t,1}^2 = - \xi m_h^2<0,\quad k_2^2 = k_{t,2}^2 = - \bar\xi m_h^2<0 \quad
2 p_1\cdot p_2 = \hat s, \nonumber\\& k_{t,1}\cdot k_{t,2} = - \sqrt{\xi \bar \xi} m_h^2 \cos\theta\nonumber.
\end{align}

The coefficient  function
$C_{\pt}(N,\xi,\bar\xi,\xi_p,\{y_i\})$ is then defined as the Mellin transform
\beq\label{eq:cmellin}
C_{\pt}(N,\xi,\bar\xi,\xi_p,\{y_i\}) = 
\int_0^1 d x x^{N-1} C_{\pt}(w,\xi,\bar\xi,\xi_p,\{y_i\})
\eeq
where
\beq \label{eq:melvar}
 x = \frac{m_h^2}{\hat s z \bar z}.
\eeq
Note that Mellin
transformation in Eq.~(\ref{eq:cmellin}) is performed for simplicity
with respect to 
the standard $\pt$-independent scaling variable Eq.~(\ref{eq:melvar})
as in the inclusive computation of Ref.~\cite{Marzani:2008az}: as
already mentioned,
computing the Mellin transform with respect to the variable $x^\prime$
Eq.~(\ref{eq:xip}) would lead to a result which differs by subleading
terms, and thus to the same final LL$x$ answer. 

The quantity  $C_{\pt}(x,\xi,\bar\xi,\xi_p,\{y_i\})$ in Eq.~(\ref{eq:cmellin})
is the transverse momentum distribution
\begin{align}
\label{eq:definc}
C_{\pt}( x,\xi,\bar\xi,\xi_p,\{y_i\}) &= \int \frac{1}{2 \hat s z\bar z}\times
\left[ \frac{1}{256} \sum_{\rm col,pol} |M(g^* g^*\to H)|^2 \right]\times\nonumber\\
\times & d \mathcal P(k_1+k_2\to p_h)\times
\delta\left(\xi_p - \xi - \bar \xi - 2 \sqrt{\xi \bar\xi} \cos\theta \right).
\end{align}
In Eq.~(\ref{eq:definc}) $d \mathcal P$ is the phase space factor
\beq
d\mathcal P(k_1+k_2\to p_h) = \frac{2\pi}{m_h^2} \delta\left(
\frac{1}{ x} - 1 - \xi - \bar \xi - 2 \sqrt{\xi \bar \xi} \cos\theta
\right) \frac{d\theta}{2\pi};
\eeq
the sum over off-shell gluon polarizations is performed using
\beq
\sum_{\rm \lambda} \epsilon^\mu_\lambda(k_i) \epsilon^{\nu*}_\lambda(k_i) = 
-2 \frac{k_{t,i}^{\mu}k_{t,i}^{\nu}}{k_{t,i}^2};
\eeq
and the flux factor is determined on the surface orthogonal to $p_{1,2}$.

After standard algebraic manipulations, $C_{\pt}$ can be written as
\begin{multline}
\label{eq:Cpt1}
C_{\pt}\(N,\xi,\bar{\xi},\xi_p,\{y_i\}\)=2\sigma_{0}\(\{y_i\}\)\int_0^1
dx\,x^{N-2}\int_0^{2\pi} \frac{d\theta}{2\pi}
\tilde{F}\(\xi,\bar{\xi},\xi_p,\{y_i\}\)\\ \delta\(\frac{1}{
  x}-1-\xi_p\)\delta\(\xi_p-\xi-\bar{\xi}-2\sqrt{\xi
  \bar{\xi}}\cos\theta\),
\end{multline}
where $\sigma_0$ is the LO Higgs production cross-section
\beq\label{eq:sigmazero}
\sigma_0\(\{y_i\}\)=\sigma_0^{\rm PL} \abs{\sum_{\{y_i\}} K\(y_i\)}^2,
\eeq
\begin{align}\label{eq:sigzpl}
\sigma_0^{\rm PL}&=\frac{\Gf \sqrt{2}\as^2}{576\pi};\\\label{eq:kap}
K\(y\)&=6 y\(1-\frac{1}{4}\(1-4y\)\ln^2\frac{\sqrt{1-4y}-1}{\sqrt{1-4y}+1}\).
\end{align} 
In Eq.~(\ref{eq:sigzpl}) (as well as in all the remaining Equations
in this paper) the branch cut in the logarithm should
be handled by giving $y$ a small negative imaginary part.  
The rather lengthy explicit formula for the form factor $\tilde F$
 is reported in Appendix~\ref{app:impact}, together with some
limiting cases. Note that the quark mass dependence is contained both 
in the Born cross-section $\sigma_0$ and in the form factor $\tilde{F}$.
Note also that if the exact quark mass dependence is retained, 
the form factor $\tilde F$ vanishes in the $\xi,\bar \xi \to \infty$ limit, while it approaches a constant
($\tilde F \to \cos^2\theta$) in the pointlike approximation.
This fact leads to a qualitatively different high-energy
behaviour in the two cases, which we will discuss in detail in the
next Sections.

Inserting the expression Eq.~(\ref{eq:Cpt1}) for the coefficient function
$C_{\pt}$ in the impact factor Eq.~(\ref{eq:hstart}) and using
one delta function to perform the $x$ Mellin integral we obtain
\begin{align}
\label{eq:hfinal}
&h_{\pt}\(N,M_1,M_2,\xi_p,\{y_i\}\)=\sigma_0\(\{y_i\}\) M_1 M_2 R\(M_1\) R\(M_2\)\frac{\xi_p^{M_1+M_2-1}}{\(1+\xi_p\)^N}
\int_0^\infty d\xi_1\,\xi_1^{M_1-1}\notag\\
&\int_{0}^{\infty} d\xi_2\,\xi_2^{M_2-1}\int_{-1}^{1}\frac{du}{\sqrt{1-u^2}}\,\frac{2}{\pi} F\(\xi_1,\xi_2,\xi_p,\{y_i\}\)\delta\(1-\xi_1-\xi_2-2\sqrt{\xi_1\xi_2}\,u\),
\end{align}
where we have introduced
\beq\label{eq:newdef}
u = \cos\theta,~~~ 
\xi_1 = \frac{\xi}{\xi_p} = - \frac{k_{t,1}^2}{\pt^2},~~~
\xi_2 = \frac{\bar\xi}{\xi_p} = - \frac{k_{t,2}^2}{\pt^2}
\eeq
and defined 
\beq
\label{eq:Fform1}
F\(\xi_1,\xi_2,\xi_p,\{y_i\}\)=\tilde{F}\(\xi,\bar{\xi},\xi_p,\{y_i\}\).
\eeq

We have performed several checks on Eq.~(\ref{eq:hfinal}). Using
the expressions in Appendix~\ref{app:impact} it is easy to see that in
the $y_i\to\infty$ limit Eq.~(\ref{eq:hfinal}) correctly reproduces
the pointlike result of Ref.~\cite{Forte:2015gve}.  Also, upon
integration over $\xi_p$ it reproduces the inclusive result of
Ref.~\cite{Marzani:2008az}. Finally, it is clear from
Eq.~(\ref{eq:newdef}) that the $\xi_p\to 0$ limit at fixed $\xi_{1,2}$
can be treated in the eikonal approximation. As explained in
Ref.~\cite{Forte:2015gve}, in this limit the result with full heavy
quark mass dependence must reduce to that of the effective theory, up
to a Wilson loop prefactor, i.e., the impact factor
Eq.~(\ref{eq:hfinal}) reduces to the pointlike result, up to the
replacement of the Born cross-section Eq.~(\ref{eq:sigmazero}) with
its pointlike form.  Comparing to the pointlike impact factor, as
given in Eqs.~(4.3,4.5) of Ref.~\cite{Forte:2015gve}, this implies the
consistency condition \beq \lim_{\xi_p \to
  0}F\(\xi_1,\xi_2,\xi_p,\yt\)=
\left(\frac{1-\xi_1-\xi_2}{2\sqrt{\xi_1\xi_2}}\right)^2 \eeq
which can be explicitly checked using the formulas in Appendix~\ref{app:impact}.

\subsection{Perturbative expansion}
\label{subsec:resres}

The perturbative expansion of the impact factor which leads to the
resummed result can now be obtained by performing the integrations
in Eq.~(\ref{eq:hfinal}). For the sake of extracting
the first several orders in the expansion of the cross-section in
powers of $\alpha_s$ we are interested in, we need the expansion of
the impact factor in powers of $M_i$. This task is not entirely
straightforward because of the $1/M_i$ collinear singularities coming
from the $\xi_i^{M_i-1}$ terms. Although the actual singularities are
 removed by the $M_i R(M_i)$ factorization terms, they prevent a naive
Taylor expansion in $M_i$. In Ref.~\cite{Forte:2015gve} this problem
was circumvented by analytically computing the impact factor for 
arbitrary values of $M_i$. In the present case, however, an analytic
computation does not appear viable because of the complexity of 
$F$ when the full quark mass dependence is retained. 

In order to extract the desired coefficients in the expansion of the impact
factor we then proceed as follows. First, we note that because of the
kinematics in the LL$x$ limit we cannot have collinear singularities
in both $\xi_1$ and $\xi_2$ at the same time. This is because the
transverse momentum of the two incoming off-shell gluons must exactly
balance the Higgs transverse momentum, so we cannot have
$\xi_1=\xi_2=0$ and $\xi_p\ne 0$ at the same time. This is made explicit by
the delta constraint in Eq.~\eqref{eq:hfinal}. It is then natural
to split the integration domain in two regions, one with $\xi_1>\xi_2$
and another with $\xi_2>\xi_1$. In the first one, we define $\xi_2 =
z \xi_1$ and rewrite Eq.~\eqref{eq:hfinal} as
\begin{align}
&h^I_{\pt}\(N,M_1,M_2,\xi_p,\{y_i\}\)=\sigma_0\(\{y_i\}\) M_1 M_2
R\(M_1\) R\(M_2\)\frac{\xi_p^{M_1+M_2-1}}{\(1+\xi_p\)^N} \int_0^1
dz\,z^{M_2-1}\notag\\ 
&\int_{-1}^{1} \frac{2 du}{\pi \sqrt{1-u^2}}
\int_{0}^{\infty}
d\xi_1\,\xi_1^{M_1+M_2-1}
F\(\xi_1,z \xi_1,\xi_p,\{y_i\}\)\delta\(1-\xi_1(1+2\sqrt{z}\,u + z)\),
\end{align}
where we have denoted with $h^I_{\pt}$ the contribution from this first
integration region.

We now use the delta function to perform the $\xi_1$ integration to
obtain
\begin{align}
\label{eq:hfinalI}
&h^I_{\pt}\(N,M_1,M_2,\xi_p,\{y_i\}\)=\sigma_0\(\{y_i\}\) M_1 M_2
R\(M_1\) R\(M_2\)\frac{\xi_p^{M_1+M_2-1}}{\(1+\xi_p\)^N} \int_0^1
dz\,z^{M_2-1}\notag\\ 
&\int_{-1}^{1} \frac{2 du}{\pi \sqrt{1-u^2}}
\left[\frac{1}{1+2\sqrt{z}u+z}\right]^{M_1+M_2}
F\(\frac{1}{1+2\sqrt{z}u+z},\frac{z}{1+2\sqrt{z}u+z},\xi_p,\{y_i\}\).
\end{align}
Note that in Eq.~\eqref{eq:hfinalI} the limit $\xi_1\to 0$ is harmless
and only the limit $z\to 0$ is associated with a collinear singularity.
We compute it using  the identity
\be
z^{M-1} = \frac{1}{M}\delta(z) + \sum_{j=0}^{\infty}\frac{M^{j-1}}{(j-1)!}
\left[\frac{\ln^{j-1} z}{z}\right]_+,
\ee
where the plus distribution is defined as
\be
\int_0^1 dz \left[f(z)\right]_+\, g(z) = \int_0^1 dz f(z) \left[g(z)-g(0)\right].
\ee

We then rewrite Eq.~\eqref{eq:hfinalI} as
\begin{align}
\label{eq:hfinalIexp}
&h^I_{\pt}\(N,M_1,M_2,\xi_p,\{y_i\}\)=\sigma_0\(\{y_i\}\) M_1 M_2
R\(M_1\) R\(M_2\)\frac{\xi_p^{M_1+M_2-1}}{\(1+\xi_p\)^N} 
\int_{-1}^{1} \frac{2 du}{\pi \sqrt{1-u^2}}\times\notag\\ 
&\times
\left(
\frac{1}{M_2} F\(1,0,\{y_i\}\)+
\int_0^1 dz
\frac{a^{M_1+M_2} F\(a,b,\xi_p,\{y_i\}\)-F\(1,0,\xi_p,\{y_i\}\)}{z} z^{M_2}
\right)
\end{align}
where we have introduced the notation 
\beq
\label{eq:abdef}
 a = a(z,u) = \frac{1}{1+2\sqrt{z}u +z},~~~ b = b(z,u) = \frac{z}{1+2\sqrt{z}u + z}.  
\eeq
In
Eq.~\eqref{eq:hfinalIexp} the collinear pole in $M_2=0$ has been 
isolated explicitly, 
and the remainder can be Taylor-expanded in $M_i$;
Eq.~\eqref{eq:hfinalIexp} only involves integrals over compact
regions, which can be easily performed numerically. Since
$F$ is symmetric under $\xi_1\leftrightarrow \xi_2$ exchange, the result
for the second region $\xi_1<\xi_2$ can now be obtained from the left hand
side of Eq.~\eqref{eq:hfinalIexp} via $M_1\leftrightarrow M_2$ exchange.

Combining the contributions from the two regions, we find that the
expansion of the impact factor
Eq.~\eqref{eq:hfinal} has the general structure
\begin{align}
\label{eq:hexpansion2}
h_{\pt}&\(N, M_1,M_2,\xi_p,\{y_i\}\)=\sigma_0\(\{y_i\}\)R\(M_1\)R\(M_2\)
\frac{\xi_p^{M_1+M_2-1}}{\(1+\xi_p\)^N}\notag \times\\ 
&\times \left[c_0\(\xi_p,\{y_i\}\)\(M_1+M_2\)+\sum_{j\ge k>0}
  c_{j,k}\(\xi_p,\{y_i\}\) \(M_1^k M_2^j+M_1^j M_2^k\)\right]
\end{align}
with
\begin{align}
\label{eq:cjk}
c_0 &\(\xi_p,\{y_i\}\)=\int_{-1}^{1}\frac{2du}{\pi\sqrt{1-u^2}} 
F\(0,1,\xi_p,\{y_i\}\)\\ 
c_{j,k}&\(\xi_p,\{y_i\}\)=\frac{1}{\(j-1\)!\(k-1\)!}\frac{1}{1+\delta_{jk}}\times\notag\\ &\int_{-1}^{1}\frac{2du}{\pi\sqrt{1-u^2}}
 \int_0^1 dz
\frac{\ln^{j-1}a\, \ln^{k-1} b\, F\(a,b,\xi_p,\{y_i\}\)
-\delta_{j,1} \ln^{k-1} z\,F\(1,0,\xi_p,\{y_i\}\)}{z} \notag \\&
+(j\leftrightarrow k)
\end{align}
and $a,b$ defined in Eq.~\eqref{eq:abdef}. A relatively simple analytic expression
for $c_0$ is presented in Appendix~\ref{app:impact}, see Eqs.~(\ref{eq:F10},\ref{eq:A1xp0}).

The expansion and resummation of the transverse momentum distribution
in the $\overline{\rm MS}$ scheme
are finally obtained by substituting the expansion
Eq.~\eqref{eq:hexpansion2} of the impact factor in
Eqs.~(\ref{eq:sigmaresxp},\ref{eq:hstart},\ref{eq:hqq}), and then letting~\cite{Jaroszewicz:1982gr,Altarelli:1999vw}
\beq\label{eq:mdual}
M_1=M_2=\gamma\(\frac{\as}{N}\)=\frac{\Ca}{\pi}\frac{\as}{N}+\Ord\(\as^4\)
\eeq
and 
\beq\label{eq:rdef}
R_{\overline{\rm MS}}\(\frac{\as}{N}\)=1+\Ord\(\as^3\).
\eeq
Note that this means that
at $\Ord\(\as\)$ (LO), only the coefficient $c_0\(\xi_p,\{y_i\}\)$
contributes to the transverse momentum distribution while 
at   $\Ord\(\as^2\)$ (NLO) we must also include 
$c_{1,1}\(\xi_p,\{y_i\}\)$,  and at $\Ord\(\as^3\)$ (NNLO)
$c_{2,1}\(\xi_p,\{y_i\}\)$ (here and henceforth we count
powers of $\alpha_s$ not including the overall $\alpha_s^2$ factor
from $\sigma_0$).

In view of our main  goal, which is to estimate finite quark mass
effect, it is interesting to compare our result
Eq.~\eqref{eq:hexpansion2} with its pointlike counterpart, as obtained in
Ref.~\cite{Forte:2015gve}. In that reference, the impact factor was
obtained in closed form:
\begin{align}
h^{\rm PL}_{\pt}&\(N,M_1,M_2,\xi_p\)=
\sigma^{\rm PL}_0 R\(M_1\)R\(M_2\)\frac{\xi_p^{M_1+M_2-1}}{\(1+\xi_p\)^N}\times\notag\\
&\times\left[\frac{\Gamma\(1+M_1\)\Gamma\(1+M_2\)\Gamma\(2-M_1-M_2\)}
{\Gamma\(2-M_1\)\Gamma\(2-M_2\)\Gamma\(M_1+M_2\)}\(1+\frac{2M_1M_2}{1-M_1-M_2}\)\right]
\end{align}
which can be expanded in power of $M_1$ and $M_2$, with the result
\begin{align}\label{eq:ifpt}
h^{\rm PL}_{\pt}\(N,M_1,M_2,\xi_p\)&=
\sigma^{\rm PL}_0 R\(M_1\)R\(M_2\)\frac{\xi_p^{M_1+M_2-1}}{\(1+\xi_p\)^N}\times\notag\\
&\times\left[c^{\rm PL}_0\(M_1+M_2\)+\sum_{j\ge k>0}c_{j,k}^{\rm PL}\(M_1^j M_2^k+M_1^k M_2^j\)\right].
\end{align}

Although Eq.~\eqref{eq:hexpansion2} and Eq.~\eqref{eq:ifpt} have the same
formal structure, if the exact quark mass dependence is retained, 
the coefficients $c_{jk}$ Eq.~\eqref{eq:cjk}
depend non-trivially on $\xi_p$, while in the pointlike approximation
they are just numbers.
The $\pt$ independence of the coefficients $c_{j,k}^{\rm PL}$ is a
reflection of the collinear origin of high-energy radiation and of the
pointlike nature of the interaction, see~\cite{Forte:2015gve}.
Nevertheless, as we already mentioned, in the $\xi_p\to 0$ limit the
pointlike result should be recovered up to an overall rescaling.  This
in particular implies that 
\beq\label{eq:ptlimit}
c_{j,k}\(\xi_p,\{y_i\}\)\underset{\xi_p \to 0}{\to} c_{j,k}^{\rm PL}.
\eeq 
Using Eq.~\eqref{eq:cjk} and the explicit form of $F$ in the $\xi_p\to 0$ 
limit given in Appendix~\ref{app:impact}, it is indeed straightforward 
to show that Eq~\eqref{eq:ptlimit} numerically holds for arbitrary $j,k$.
The situation is rather different in the opposite $\xi_p\to\infty$
limit.  Indeed, in this case it is clear from Eq.~\eqref{eq:ifpt} that
the pointlike impact factor behaves like $h_{\pt}^{\rm PL}\sim \ln^j
\xi_p/\xi_p$. On the other hand, thanks to the presence of the form
factor $F$ in Eq.~\eqref{eq:cjk} the impact factor in the full theory
vanishes at least as $h_{\pt}\sim \ln^j\xi_p/\xi_p^2$, leading to a much
softer high $\pt$ spectrum. 

\section{Parton-level Results}
\label{sec:parton}

We now present and discuss results for the partonic
cross-section in the gluon channel obtained from the expansion
Eq.~(\ref{eq:hexpansion2}-\ref{eq:cjk}) of the resummed
results. 
We will specifically include top and bottom masses, i.e. henceforth
$\{y_i\}=\{\yt,\yb\}$. 
We expect the coefficients
$c_{j,k}\(\xi_p,\{y_i\}\)=c_{j,k}\(\xi_p,\yt,\yb\)$ to depart from the 
pointlike limit when the transverse momentum
starts resolving the top loop, for $\xi_p\sim y_t$, 
and also to show some smaller
deviation from the pointlike behaviour in the region $\xi_p \gsim y_b$
in which the bottom mass effects are felt.

First, we compare the exact result, which as mentioned is only known
at LO, to our high-energy result, and to the pointlike limit. Then, we
discuss the structure of the first several perturbative expansion
coefficients $c_{i,j}$ Eq.~(\ref{eq:cjk}), and specifically compare the
pointlike limit to the contributions of top, bottom and interference.
Finally, we use our result to address the issue the possible
exponentiation of bottom logs in the intermediate scale region $m_b<
\pt <m_H$ which has been the object of some recent
discussion~\cite{Bagnaschi:2011tu,Banfi:2013eda,Grazzini:2013mca,Bagnaschi:2015qta,Melnikov:2016emg}.

Here and in the rest of this paper we will show all results for 
$m_h=125.09~{\rm GeV}$ and with
heavy quark masses given as pole masses, with the values
\begin{align}
m_t&=173.07\,\rm GeV, & \yt&=1.914; \\
m_b&=4.179\,\rm GeV, & \yb&=0.00112.
\end{align}
Note that the difference between pole and $\overline{\rm MS}$ masses
is NLL$x$, and thus for our LL$x$ results only the numerical value of
the heavy quark mass matters. Similarly, different scale choices only
affect our predictions at NLL$x$. As explained in
Sect.~\ref{sec:resummation}, we set $\mur=Q$ Eq.~\eqref{eq:physcale}
for the numerical results shown in this Section.

\subsection{Leading order: comparison to the exact result} 
\label{subsec:HEVSLO}

At leading $\Ord(\alpha_s)$ our result reduces to \beq\label{eq:lohe}
\frac{d\hat\sigma^{\text{LL}x-\text{LO}}}{d\xi_p}=
\sigma_0\(\yb,\yt\)c_0\(\xi_p,\yt,\yb\)\frac{2\Ca\as}{\pi}
\frac{1}{\xi_p}, \eeq with $\sigma_0$ given by
Eqs.~(\ref{eq:sigmazero},\ref{eq:kap}) and $c_0$, Eq.~\eqref{eq:cjk};
note in particular that it does not depend on $x$ because the LL$x$
cross section is proportional to $\sigma_0 \,\as^k \ln^{k-1}x$,
$k>0$. The coefficient $c_0$ can be determined in fully analytic form,
see Eqs.~(\ref{eq:F10},\ref{eq:A1xp0}).

\begin{figure}[htb!]
\centering
\includegraphics[width=0.48\textwidth]{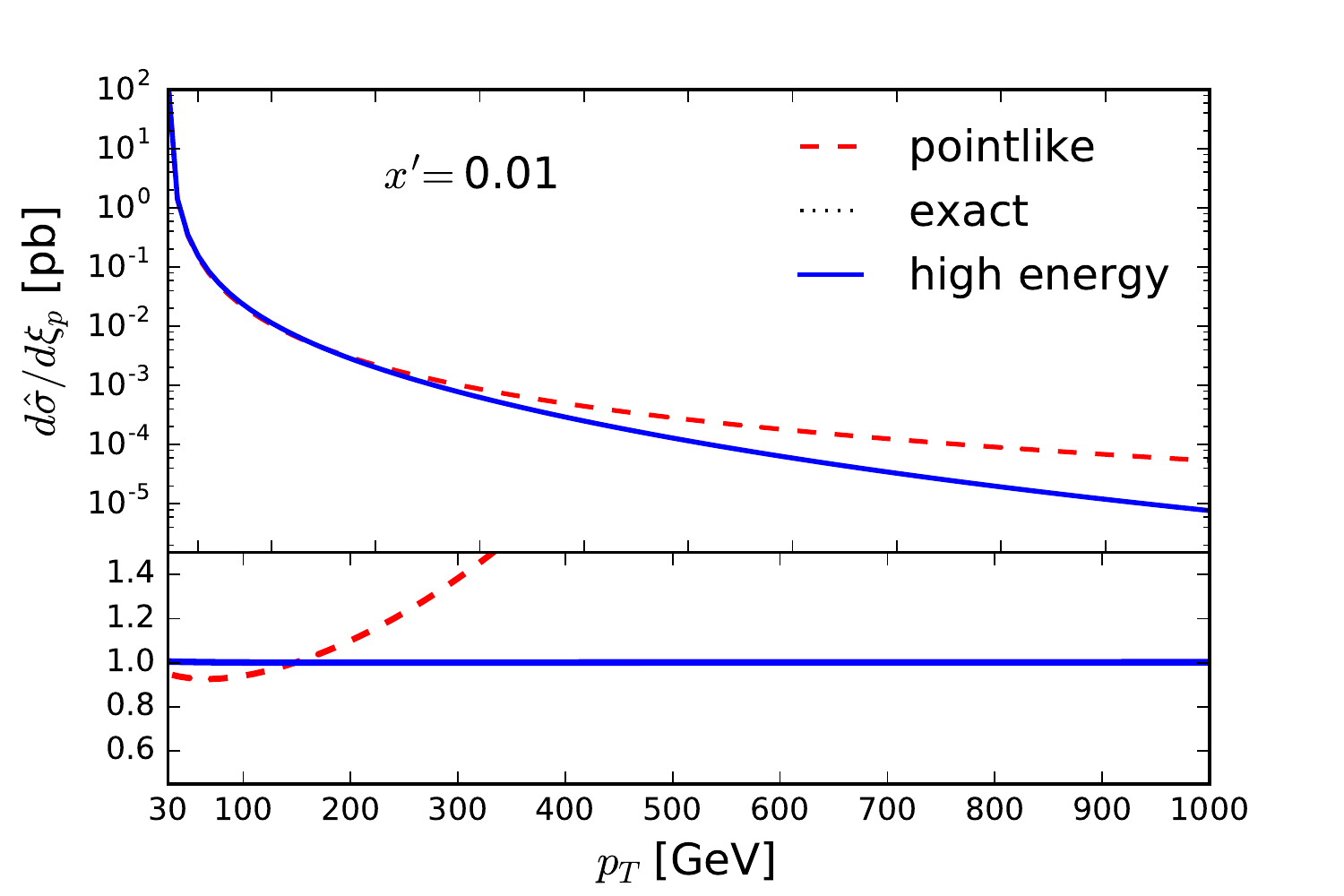}
\includegraphics[width=0.48\textwidth]{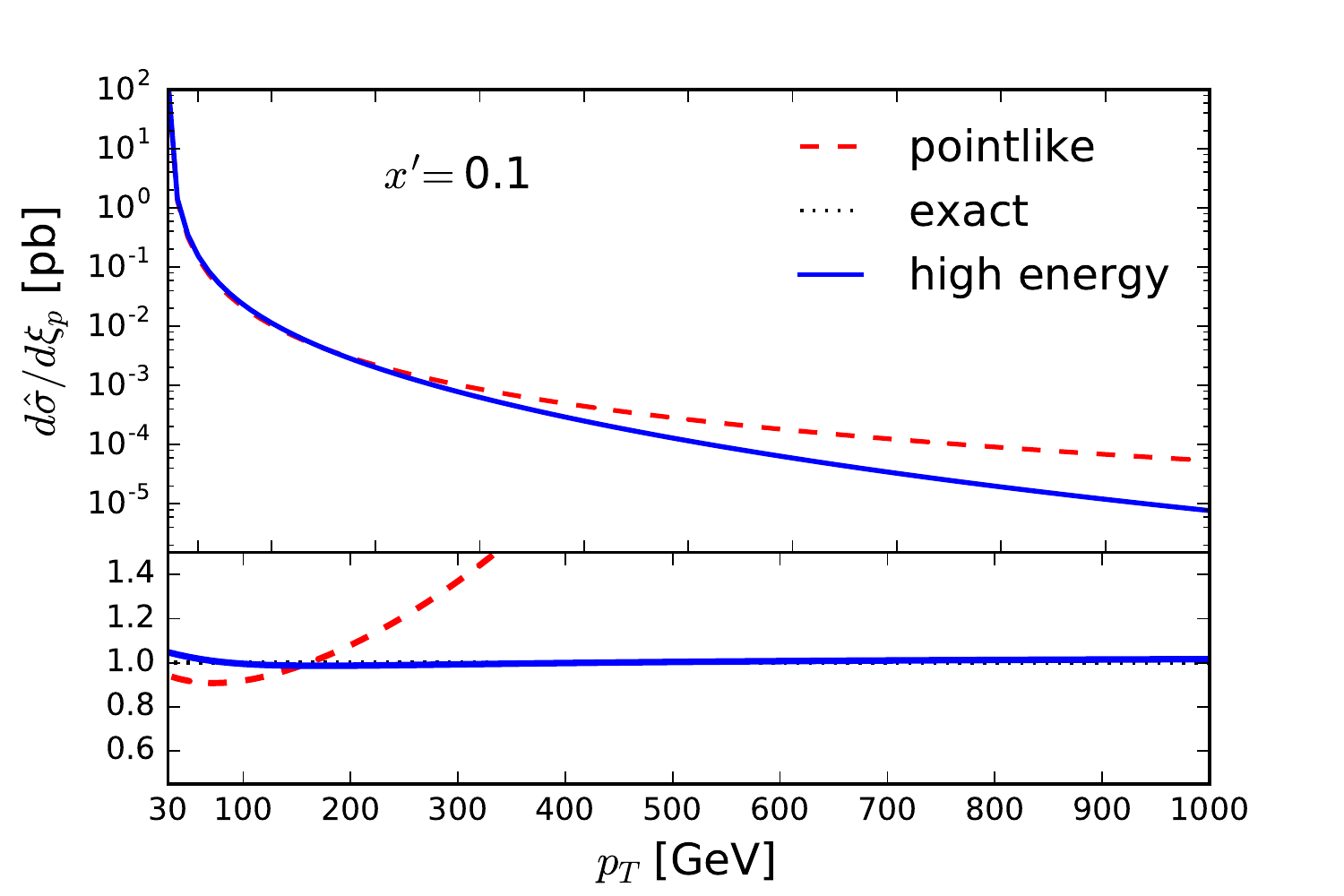}\\
\includegraphics[width=0.48\textwidth]{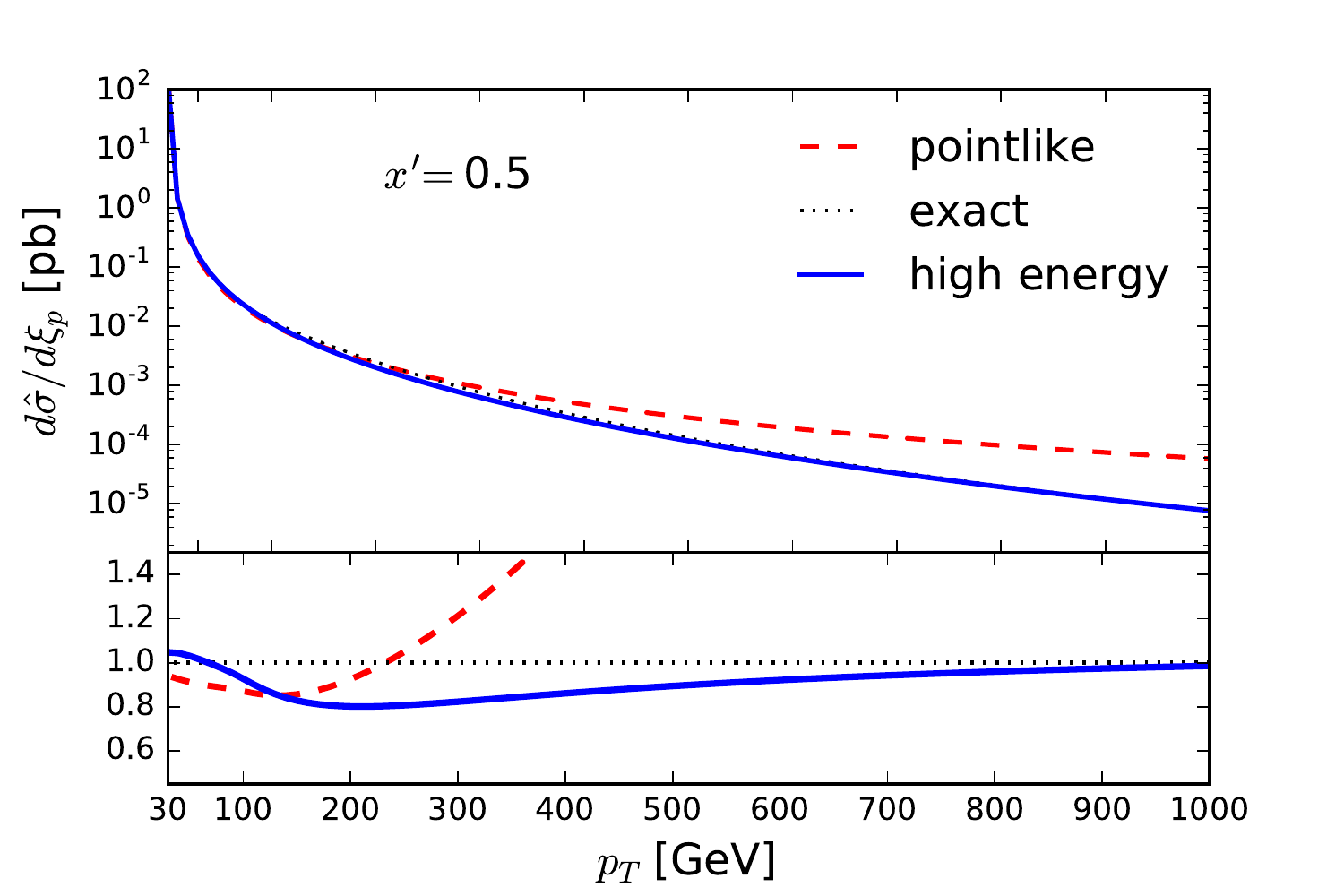}
\includegraphics[width=0.48\textwidth]{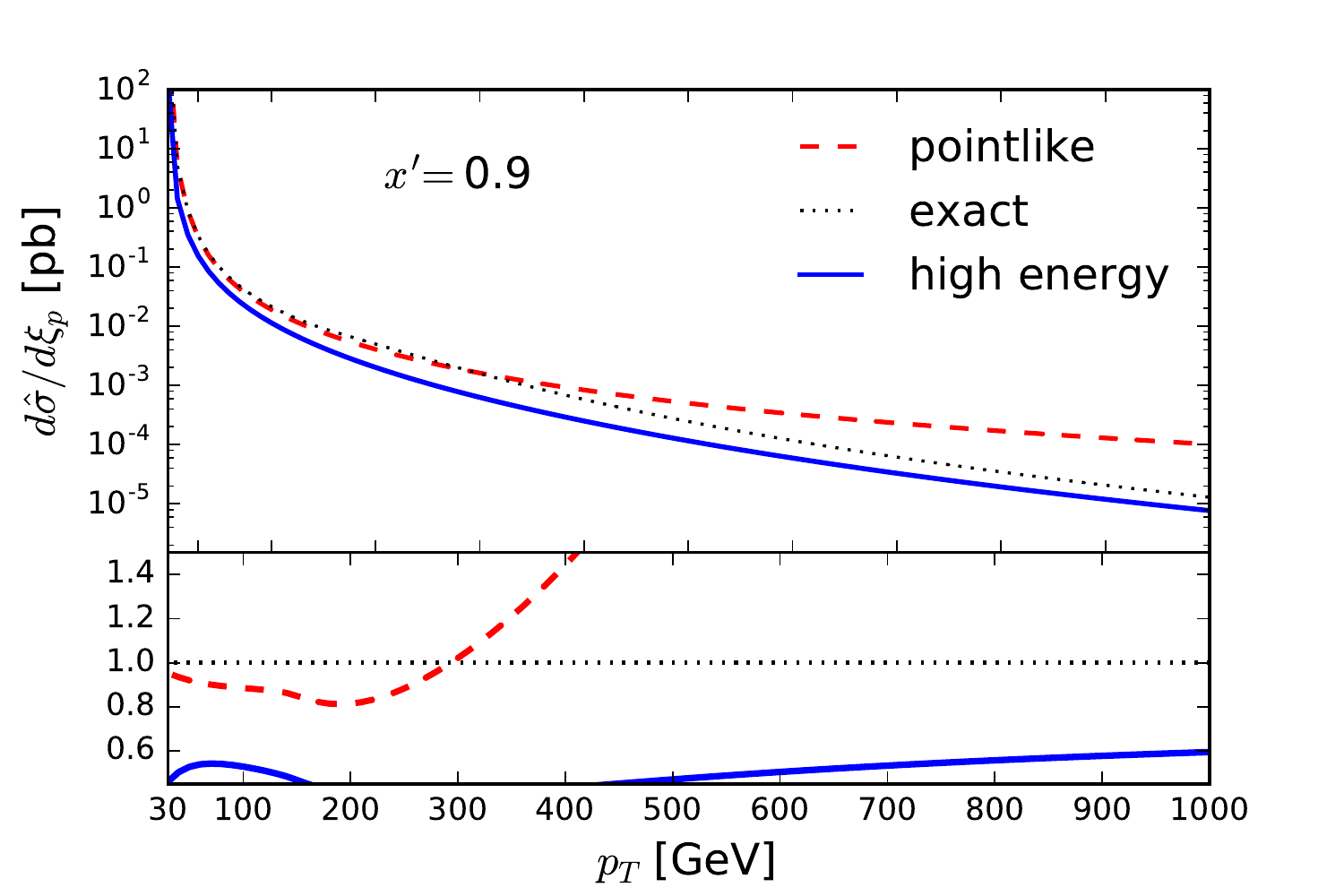}
\caption{The 
  partonic leading order transverse momentum 
distribution  in the 
  high-energy limit (blue, solid) compared to the exact result
  of Ref.~\cite{Baur:1989cm} (black, dotted). The leading order pointlike
  result~\cite{Ellis:1987xu} is also shown for comparison (red,
  dot-dashed). Results are shown for three different values of
  $x'$ Eq.~(\ref{eq:xip}): $x'=0.01$ (top left), $x'=0.1$ (top right),
$x'=0.5$ (bottom left), $x'=0.9$ (bottom right);
  in each case, the ratio to the exact result is also
  plotted.}
\label{fig:checkwithBaur}
\end{figure}

In Figure~\ref{fig:checkwithBaur} we compare the
exact~\cite{Baur:1989cm}, high-energy and
pointlike~\cite{Ellis:1987xu}  LO results
for four different values of $x'$ Eq.~(\ref{eq:xip}). Here and
henceforth we only show predictions for large enough $p_T>30$~GeV: at
lower $p_T$ fixed-order predictions cease to be valid, and must be
improved through Sudakov resummation. The relation of the latter to
the high-energy approximation was recently discussed in
Ref.~\cite{Marzani:2015oyb}. 
As expected, the pointlike approximation breaks down for $\pt\gsim m_t$ 
where the finite-mass result drops rather faster; the deficit
which is seen in the pointlike result for $\pt< m_t$ is due to the
finite bottom mass. The high-energy approximation appears to be very
accurate for $x'\lsim 0.1$; for higher $x'$ values it starts
deteriorating and for large $x'\sim 0.5$ it is typically off by
20\%. However, the accuracy of the high-energy approximation does not
depend on $\pt$ if $\pt\gsim m_H$: the large-$\pt$ behaviour of the high-energy
approximation is qualitatively the same as that of the full
result.

The pointlike approximation instead departs from the
exact result by an increasingly large amount as $\pt$ grows: in fact,
as $\pt\to\infty$, $c_0\(\xi_p,\yt,\yb\)$ Eq.~(\ref{eq:lohe}) drops at
least as $\frac{1}{\pt^2}$, while it is constant in the effective
theory, so $\frac{d\sigma^{\text{LL}x-\text{LO}}}{d\xi_p}
\twiddlesinf{\pt}\frac{1}{\(\pt^2\)^a}$  with $a=2$ in the full theory, and
$a=1$ in the effective theory. In the opposite limit $\pt\to 0$ instead,
as discussed in Section~\ref{subsec:resres} (see Eq.~(\ref{eq:ptlimit})),
the high-energy limit becomes pointlike, up to an
overall rescaling: it is indeed clear from the plots that 
 in the region $x'\lsim 0.1$ in which the high-energy
approximation holds, as $\pt\to0$ the high-energy
and pointlike results coincide.  An immediate consequence of this
discussion is that in the large $\pt\gsim m_t$ region it is generally
rather more advantageous to rely on the high-energy approximation,
than use pure effective field theory results, as we will discuss in more
detail in Section.~\ref{sec:hadro}.

\subsection{Expansion coefficients beyond the leading order}
\label{subsec:comparison}

We now study the expansion coefficients of the impact factor
Eq.~\eqref{eq:hexpansion2}, which we compute including both top and
bottom mass, i.e. using Eq.~(\ref{eq:cjk}). As discussed in the end of
Sect.~\ref{subsec:resres}, the LL$x$ transverse momentum distribution
up to NNLO is fully determined from knowledge of the first three
coefficients. Explicitly, using  Eq.~\eqref{eq:hexpansion2} with
Eqs.~(\ref{eq:mdual}-\ref{eq:rdef})  and inverting the Mellin
transform Eq.~(\ref{eq:melpt}) we get
\beq
\label{eq:sigmapartexp}
\frac{d\sigma}{d\xi_p}\(x,\xi_p,\yt,\yb\)=\sigma_0\(\yt,\yb\)\sum_{k=1}^\infty C_k\(\xi_p,\yt,\yb\)\as^k (-1)^{k+1} \frac{\ln^{k-1} x}{(k-1)!}
\eeq
with 
\begin{subequations} 
\label{eq:coefffulltheory}
\begin{align}
C_1\(\xi_p,\yt,\yb\)&=\frac{2\Ca}{\pi} \frac{c_0\(\xi_p,\yt,\yb\)}{\xi_p}\\
C_2\(\xi_p,\yt,\yb\)&=\frac{2\Ca^2}{\pi^2} \frac{2 c_0\(\xi_p,\yt,\yb\)\ln\xi_p+c_{1,1}\(\xi_p,\yt,\yb\)}{\xi_p}\\
C_3\(\xi_p,\yt,\yb\)&=\frac{2\Ca^3}{\pi^3} \frac{2 c_0\(\xi_p,\yt,\yb\)\ln^2\xi_p+2 c_{1,1}\(\xi_p,\yt,\yb\)\ln\xi_p+c_{2,1}\(\xi_p,\yt,\yb\)}{\xi_p}.
\end{align}
\end{subequations}
Note that the leading power of $\ln\xi_p$ is always proportional to the
lowest order coefficient $c_0$.

\begin{figure}[htb!]
\centering
\includegraphics[width=0.6\textwidth]{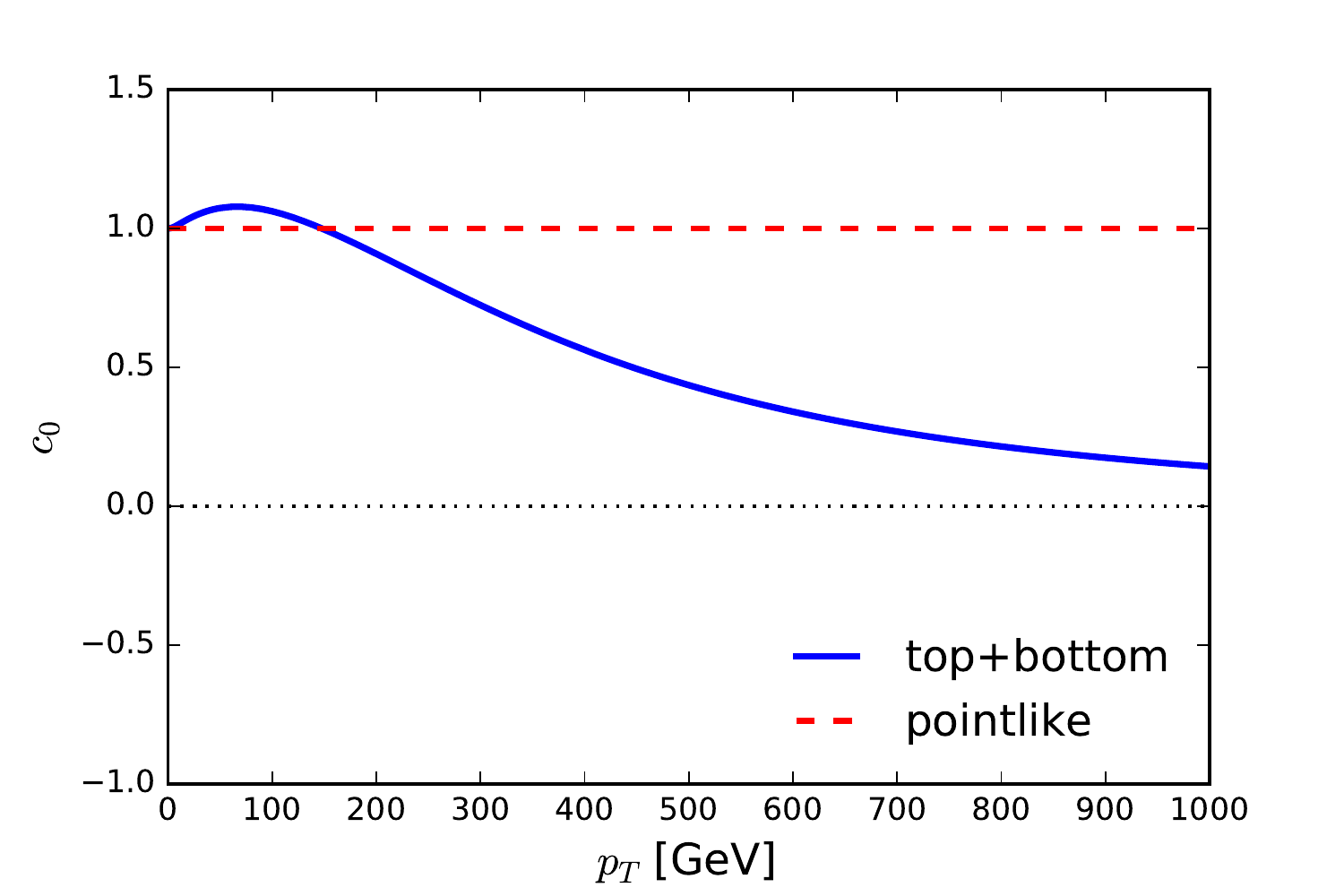}
\includegraphics[width=0.6\textwidth]{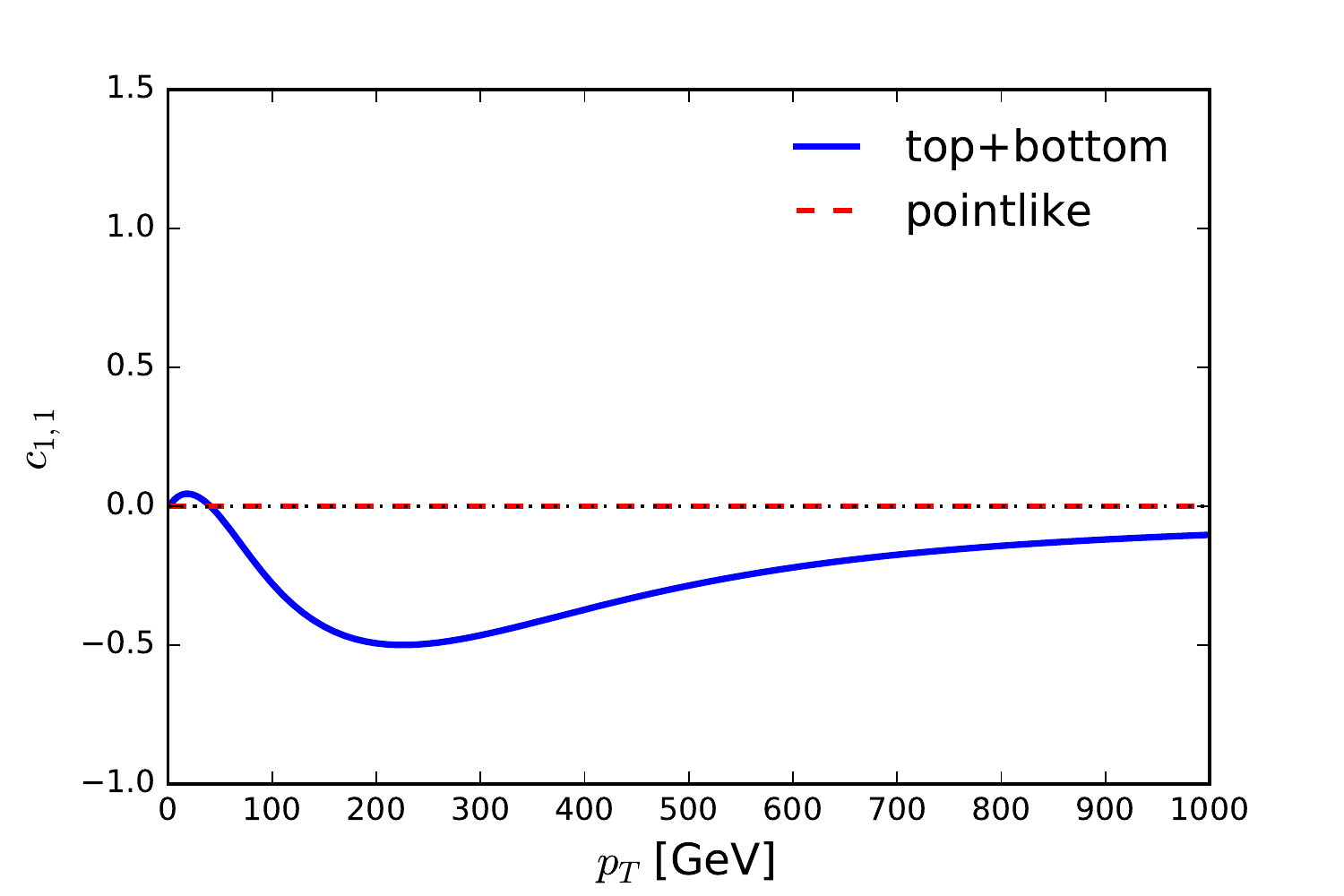}
\includegraphics[width=0.6\textwidth]{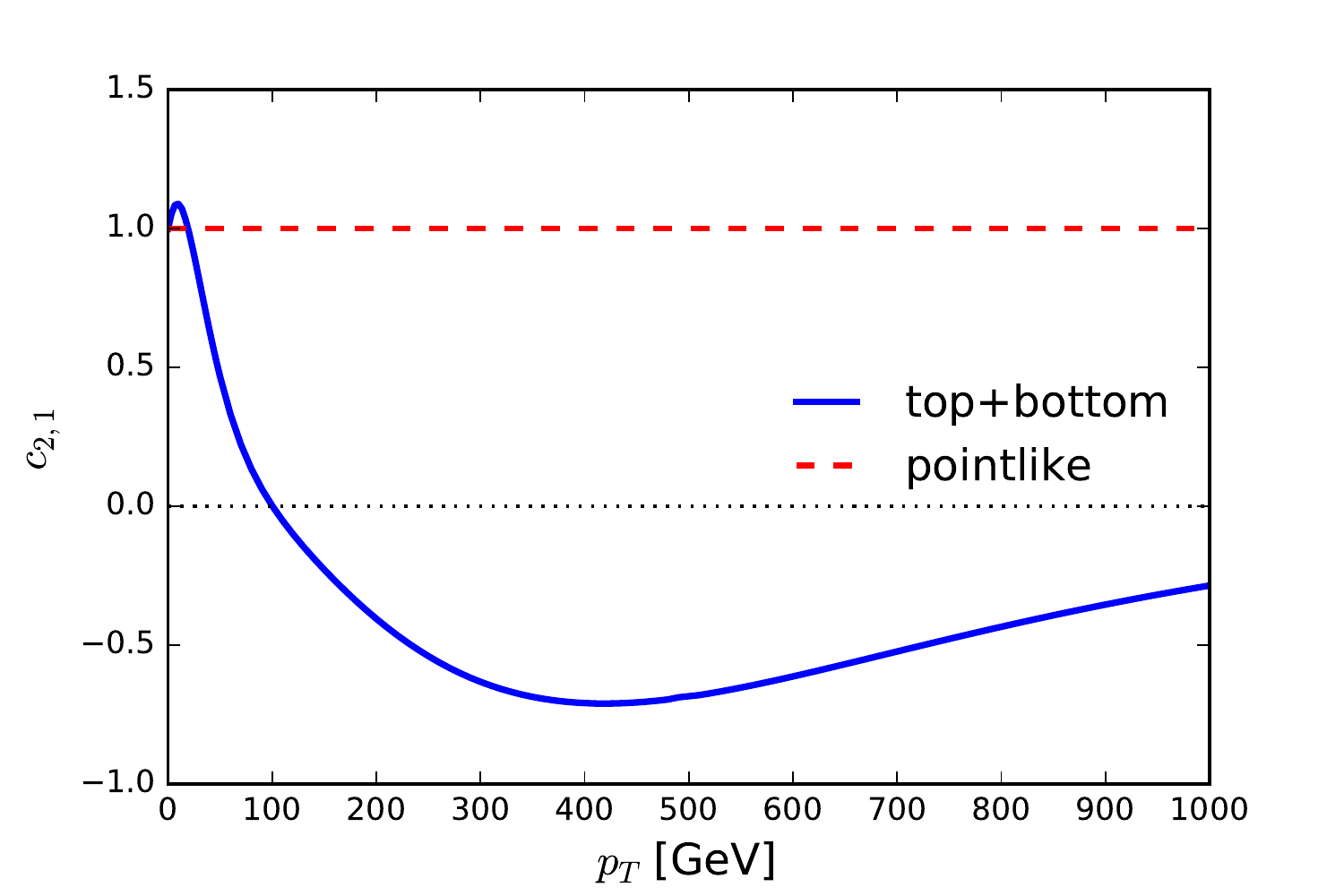}
\caption{The first three coefficients $c_{i,j}$ Eq.~\ref{eq:cjk} in
  the expansion of the transverse-momentum dependent impact factor
  Eq.~(\ref{eq:hexpansion2}) with finite top and bottom masses,
  compared to the pointlike result.}
\label{fig:coefftot}
\end{figure}

The coefficients are shown in Fig.~\ref{fig:coefftot}, and compared to
their (constant) pointlike counterparts~\cite{Forte:2015gve}. As expected, the coefficients tend to the
pointlike limit as $\xi_p\to0$, while they vanish  at large $\xi_p$,
as required in order for the inclusive cross-section to be free of
spurious double energy logs, as discussed in
Ref.~\cite{Forte:2015gve}. In the high-energy limit, the overall power 
behaviour at large $p_T$ remains the
same to all orders, and equal to that of the leading order, which as
we have seen above, coincides with theat of the exact leading
order. The fact that the high-energy approximation holds as
$x^\prime\to1$, while in the opposite
$x'\to1$ limit the high-$\pt$ power behaviour is also to all orders the
same of the leading-order result~\cite{deFlorian:2005fzc} suggests
that the high-energy approximation reproduces  the correct high-$p_T$
behaviour of the full result to all orders.

As seen in Sect.~\ref{subsec:HEVSLO}, the pointlike approximation
breaks down for $\pt\sim m_t$. In the high-energy limit, one expects  the
departure from 
pointlike to become increasingly marked as the perturbative order
is raised, because with an increasingly large  number of hard emissions
more energy flows into the loop which is less well approximated by a
pointlike interaction: so  higher-order coefficients $c_{i,j}$ deviate more from
their pointlike limit than lower-order ones. On the other hand, the
lower order coefficients are enhanced by higher powers of $\ln\xi_p$, see
 Eqs.(\ref{eq:sigmapartexp}-\ref{eq:coefffulltheory}), so
low-order coefficients dominate, and the shape of the $\pt$
distribution remains similar as the perturbative order is increased,
as we will also discuss at the hadronic level in Sect.~\ref{sec:hadro}.

Also as discussed in Sect.~\ref{subsec:HEVSLO}, the effect of the bottom
quark can be seen in the departure of the coefficients from the
pointlike value at small $\pt$, even though the pointlike limit is
always recovered in the  $\pt\to 0$ limit. 
It is interesting to assess the
relative impact
of the bottom, top, and interference contributions. In order to do
this, we write each coefficient as
\begin{equation}
\label{eq:cjktopbottom}
c_{j,k}\(\xi_p,\yt,\yb\)=R^{\rm t}\(\yt,\yb\)\:c_{j,k}^{\rm t}\(\xi_p,\yt\)
+ R^{\rm b}\(\yt,\yb\)\:c_{j,k}^{\rm b}\(\xi_p,\yb\)
+R^{\rm i}\(\yt,\yb\)\: c_{j,k}^{\rm i}\(\xi_p,\yt,\yb\),
\end{equation}
where the normalization ratios
\begin{subequations}
\label{eq:rco}
\begin{align}
R^{\rm t}\(\yt,\yb\)&=\frac{\abs{K(\yt)}^2}{\abs{K(\yt)+K(\yb)}^2}=1.107\\
R^{\rm b}\(\yt,\yb\)&=\frac{\abs{K(\yb)}^2}{\abs{K(\yt)+K(\yb)}^2}=0.008 \\
R^{\rm i}\(\yt,\yb\)&=\frac{K(\yb)^* K(\yt)+K(\yb) K(\yt)^*}{\abs{K(\yt)+K(\yb)}^2}=-0.115.
\end{align}
\end{subequations}
account for the mismatch in normalization between the Wilson
coefficients $K$ in the form factor Eq.~\eqref{eq:Fform1} when both the top and bottom
contributions are included.

\begin{figure}[hb]
\centering
\includegraphics[width=0.48\textwidth]{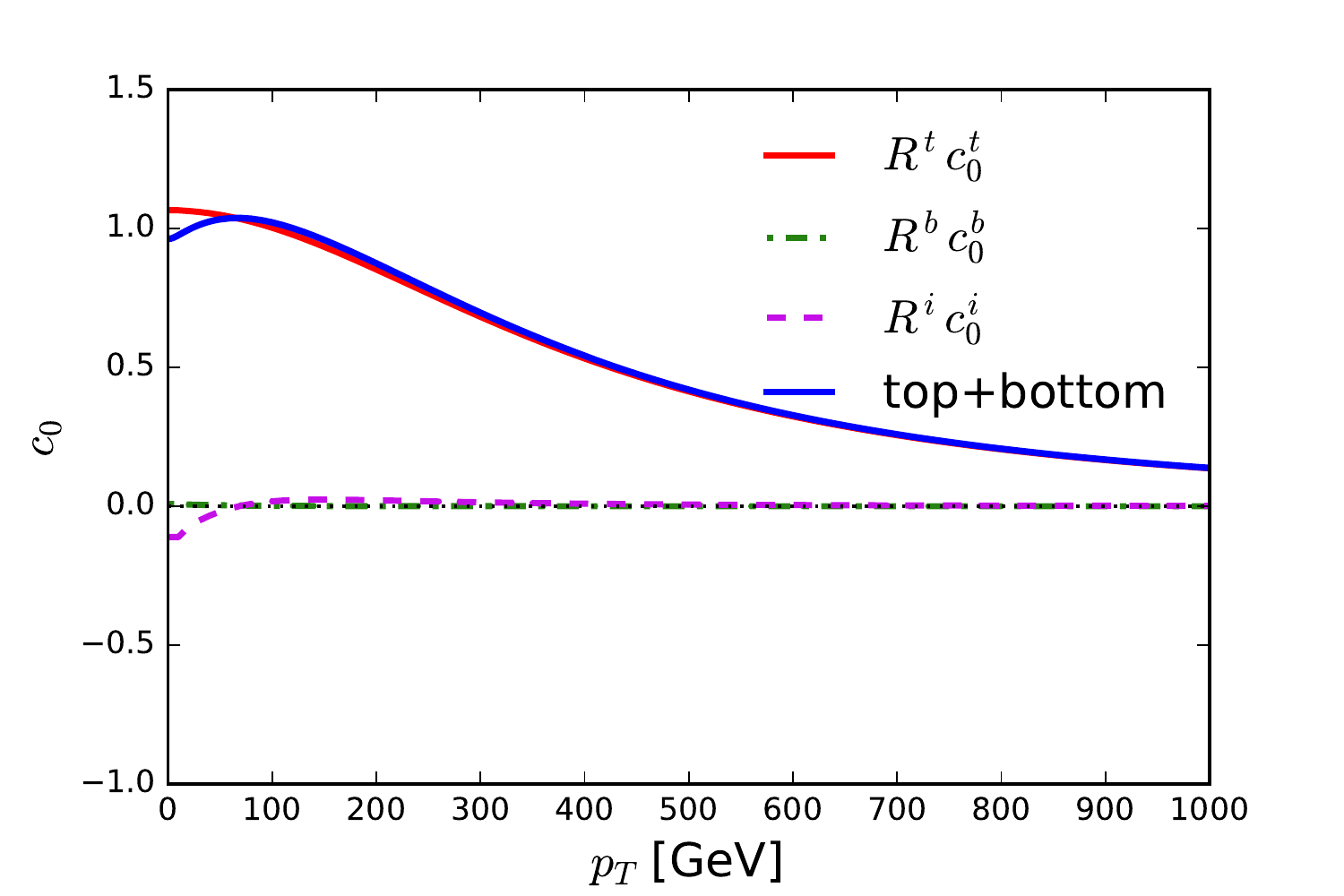}
\includegraphics[width=0.48\textwidth]{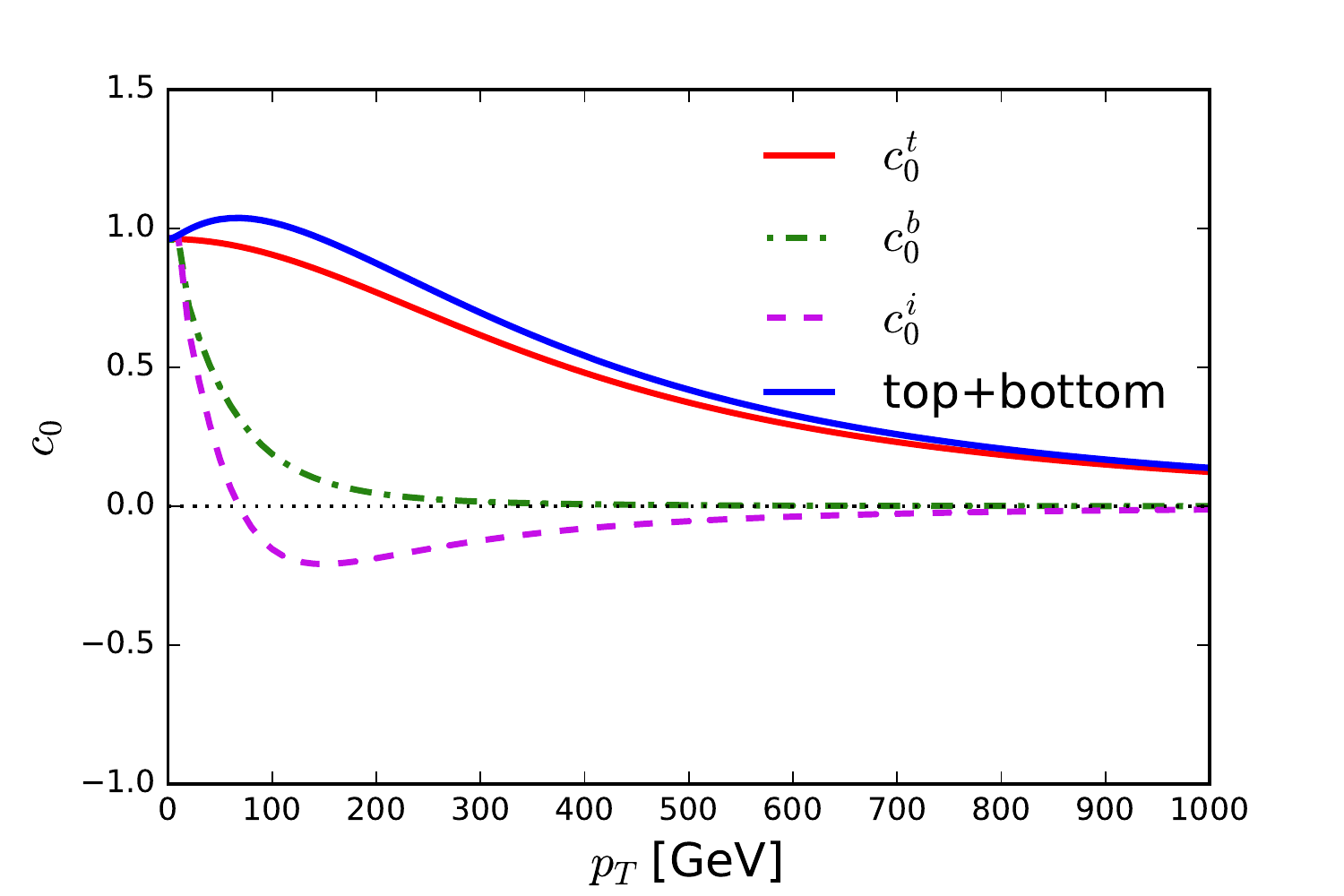}\\
\includegraphics[width=0.48\textwidth]{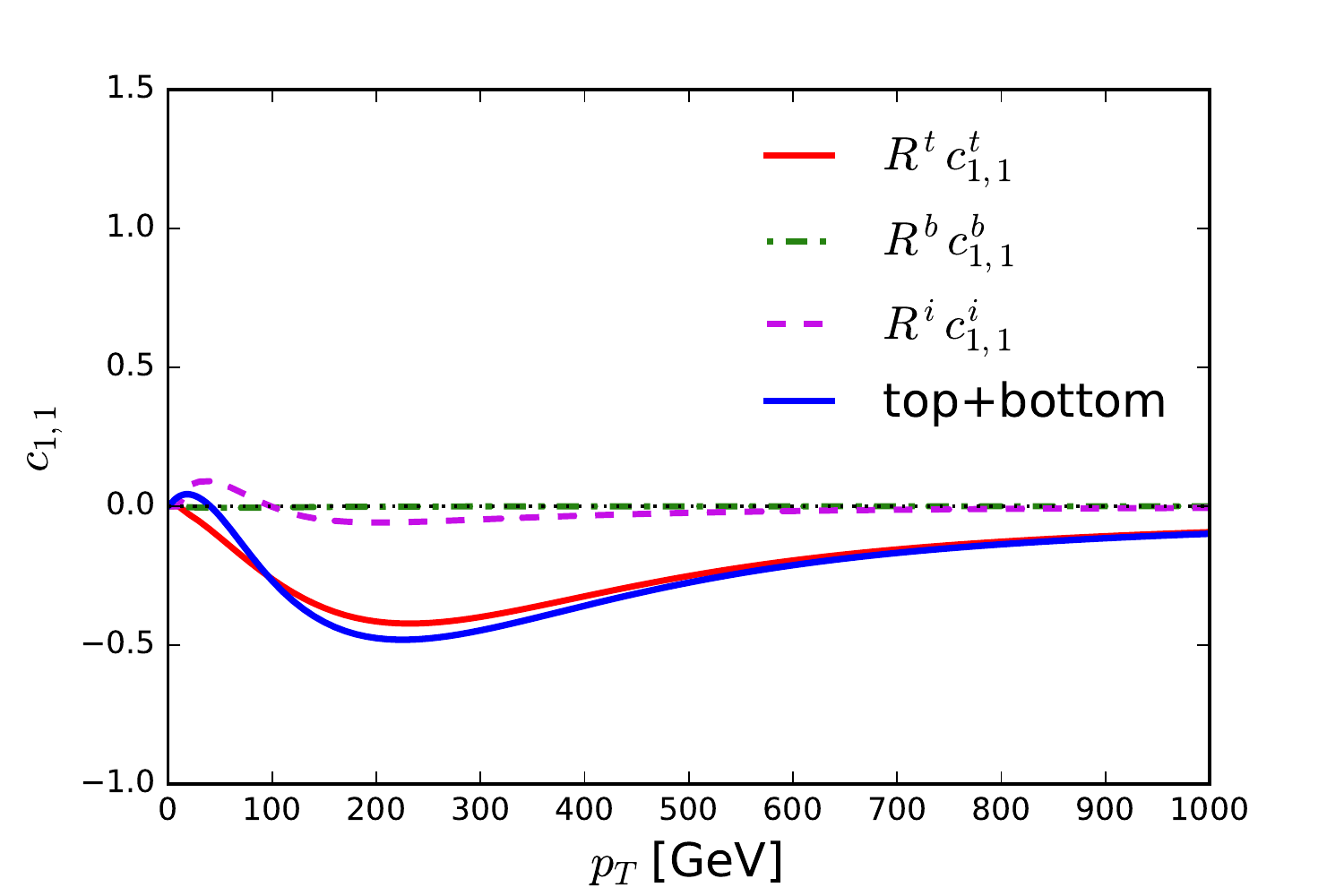}
\includegraphics[width=0.48\textwidth]{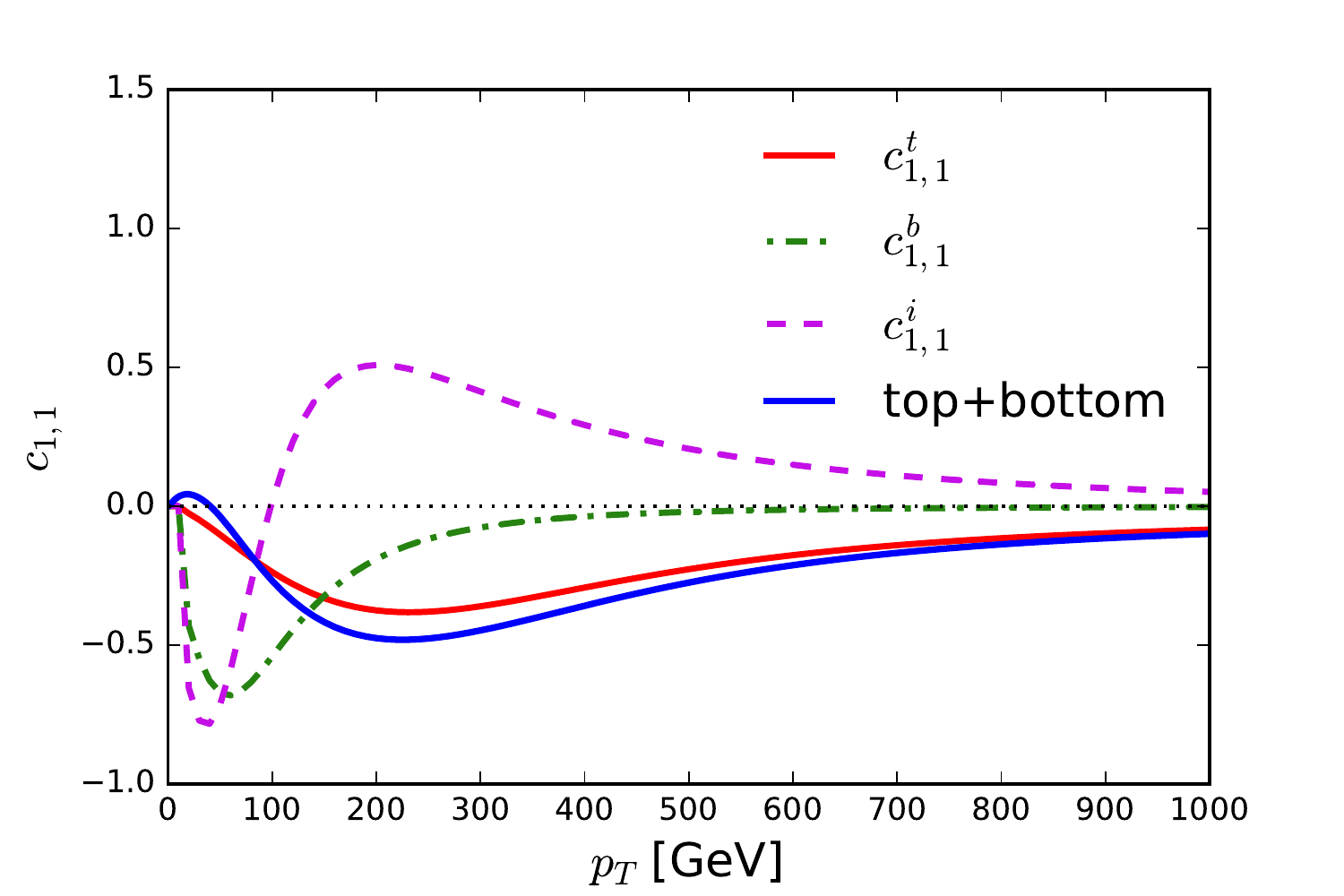}\\
\includegraphics[width=0.48\textwidth]{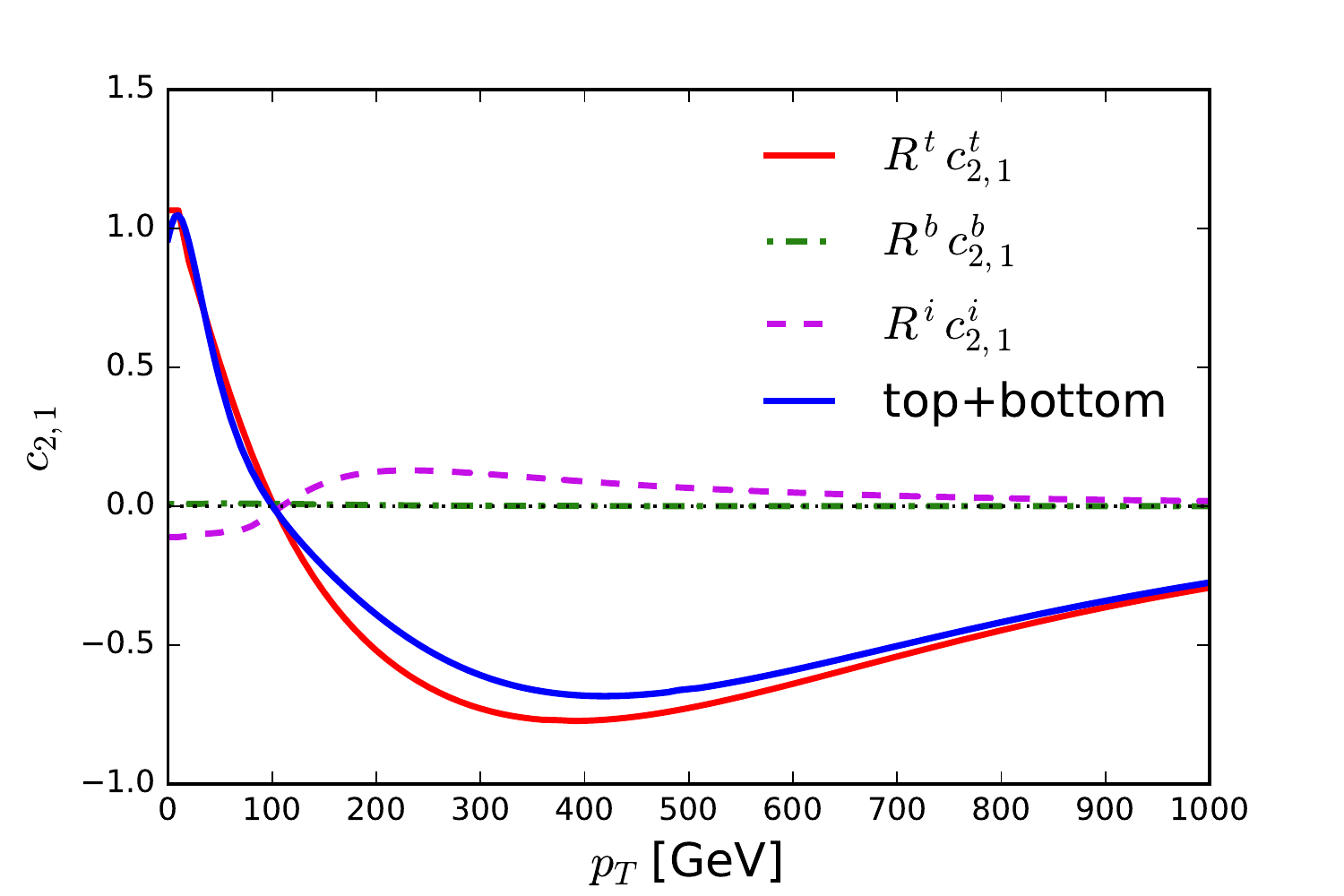}
\includegraphics[width=0.48\textwidth]{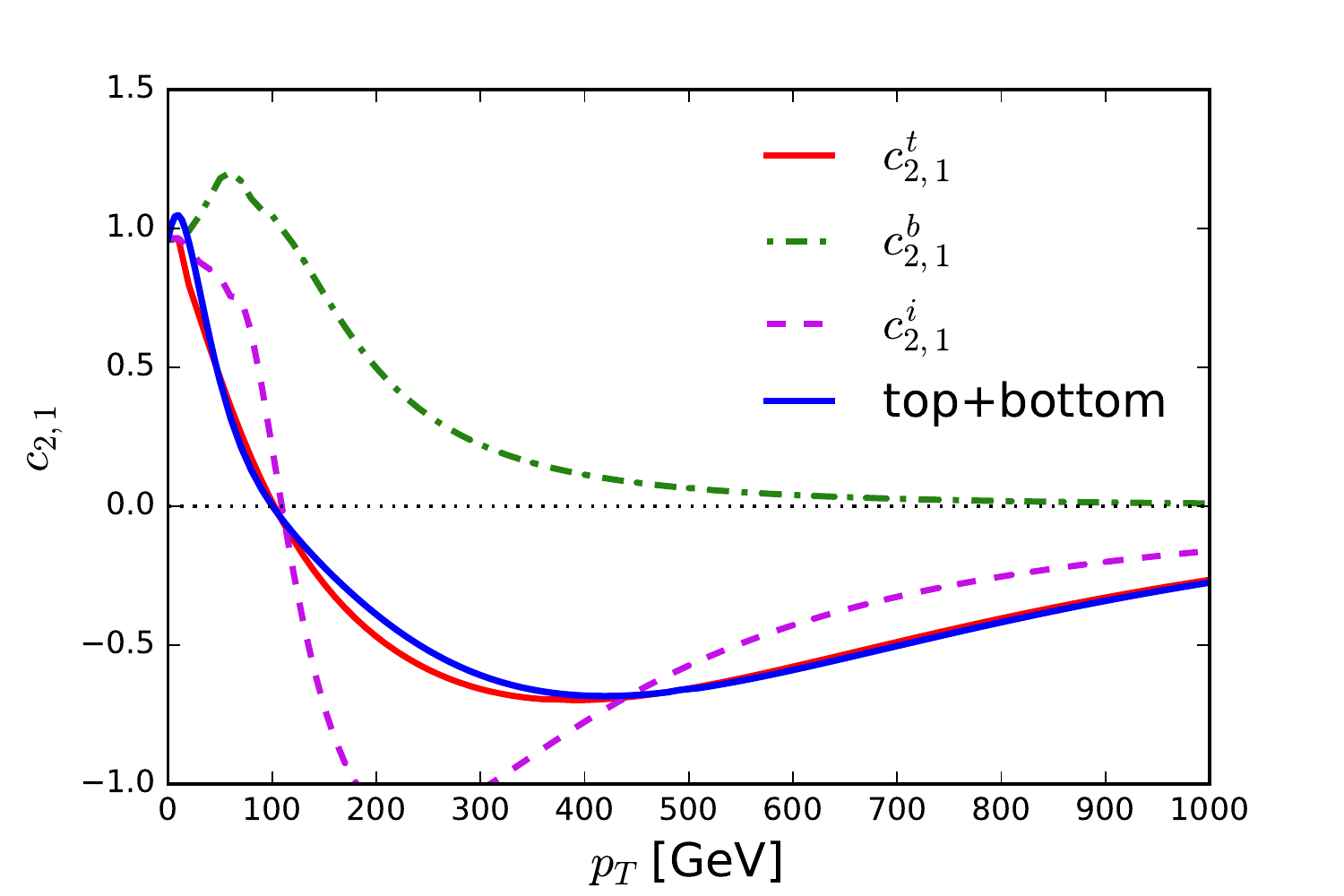}
\caption{Contribution from top (red, solid), bottom (green, dot-dashed) and interference
  (purple, dashed)  to the
  coefficients shown in Fig.~\ref{fig:coefftot}, with their sum also
  shown as blue line: the three coefficients $c_{0}$, $c_{1,1}$ and
  $c_{2,1}$ are shown from top to bottom, including (left) or not
  including (right) the normalization due to the Wilson coefficient
  Eqs.~(\ref{eq:cjktopbottom}-\ref{eq:rco}).}
\label{fig:topbottomcontrib}
\end{figure}

The separate contributions are compared in
Fig.~\ref{fig:topbottomcontrib} to each other and to their sums,
already shown in Fig.~(\ref{fig:coefftot}), both with and without the
normalization coefficients Eq.~\eqref{eq:rco}. It is clear that while
in each case the un-normalized coefficients $c^{\rm t}$, $c^{\rm b}$ and $c^{\rm i}$ are
all of the same order, after multiplying by the Wilson coefficients
Eqs.~(\ref{eq:rco}) the top contribution is dominant, while the pure
bottom contribution becomes entirely negligible. 
However, in the
region $m_b\lsim \pt\lsim m_t$ and even for somewhat larger $\pt$
values the interference contribution provides a small but
non-negligible correction. Fig.~\ref{fig:topbottomcontrib} shows that
this feature, well known at LO, appears to persist also at higher
orders. 
In this range of $\pt$, the transverse momentum spectrum acquires a
dependence on $\ln \frac{\pt^2}{m_b^2}$, as we now discuss.

\subsection{Bottom Logs}
\label{subsec:bottomlogs}

The region in which $m_b\lsim \pt\lsim m_t$ is particularly intricate
because the Higgs momentum spectrum becomes a multi-scale
problem. Indeed, it was pointed out in Ref.~\cite{Bagnaschi:2011tu}
that in this region finite bottom mass effect are visible in the
spectrum, which thus deviates from the prediction obtained using
transverse momentum resummation. Specifically, in
Ref.~\cite{Banfi:2013eda} it was shown  that  the
cross-section contains contributions proportional to $\ln
\frac{\pt^2}{m_b^2}$ which can be traced to non-factorized soft or
collinear logs.  This immediately raises the question whether
such behaviour persists at higher orders, perhaps requiring
resummation~\cite{Banfi:2013eda,Grazzini:2013mca,Bagnaschi:2015qta}. 
The resummation of these soft logs as recently discussed in
Ref.~\cite{Melnikov:2016emg}; in the high energy limit considered here
 we focus on the collinear ones instead.

It turns out in fact that, in the high-energy limit, collinear 
logs are present to all perturbative orders, but not of increasingly
high logarithmic order,  
at least at the LL$x$ level.
To see this, we first consider our LO result Eq.~(\ref{eq:lohe}) when
$m_b^2 < \pt^2 < m_h^2$~\cite{Banfi:2013eda}, i.e., using dimensionless  variables
$\yb < \xi_p < 1$. Collinear bottom mass logs are extracted by performing the
simultaneous limit 
$\frac{y_b}{\xi_p}\to0$ and $\xi_p\to0$~\cite{Banfi:2013eda}. 
We get 
\beq\label{eq:collimit}
\frac{d\sigma^{LO} }{d\xi_p}\(x,\xi_p,\yb\)\twiddles{y_b}\frac{\Gf\sqrt{2}
  \as^2 }{256\pi^2}\frac{2\Ca\as}{\pi \xi_p}\yb^2 \abs{\ln^2\frac{\xi_p}{\yb}-\ln^2\(-\yb\)+4}^2.
\eeq
But $\frac{\xi_p}{\yb}=\frac{\pt^2}{m^2_{b}}$, so this agrees with the
conclusion of  Ref.~\cite{Banfi:2013eda} that the transverse momentum
spectrum 
contains a collinear contribution proportional  to $ \frac{m_b^4}{\pt^2 m_H^2}     \ln^4 \frac{\pt^2}{m^2_{b}}$.

The corresponding result at all orders can be obtained by performing
the same limit on the function $F$ Eq.~\eqref{eq:Fform1}, which
contains all the  $\xi_p$ and $\yb$ dependence of the resummed
result. We get
\begin{multline}
F\(\xi_1,\xi_2,\xi_p,\yb\)\twiddles{y_b}\Bigg[p_{4}\(\xi_1,\xi_2\) \ln^4 \frac{\xi_p}{\yb}+p_{3}\(\xi_1,\xi_2\) \ln^3 \frac{\xi_p}{\yb}\\+p_{2}\(\xi_1,\xi_2\) \ln^2 \frac{\xi_p}{\yb}+p_{1}\(\xi_1,\xi_2\) \ln \frac{\xi_p}{\yb}+p_{0}\Bigg],
\end{multline}
where the coefficient of the highest log has the simple form
\beq\label{eq:pfour}
p_4\(\xi_1,\xi_2\)=\frac{\(1-\xi_1-\xi_2\)^2}{4\xi_1\xi_2},
\eeq
and we omit the lengthy expressions of the other coefficients.
Using Eq.~(\ref{eq:pfour}) in Eq.~\eqref{eq:Fform1}  the integrals over $\xi_i$ in the
expression of the coefficients can be performed analytically, and we
find that
the leading contribution to the impact factor in the limit is 
\begin{align}
\label{eq:impactpfour}
h_{\pt}\(N,M_1,M_2,\xi_p,\yb\) \twiddles{y_b} & \sigma^{\rm PL}_0 R\(M_1\)R\(M_2\)\frac{\xi_p^{M_1+M_2-1}}{\(1+\xi_p\)^N}\notag\\
&\left[c^{\rm PL}_0\(M_1+M_2\)+\sum_{j>k>0}c_{j,k}^{\rm PL}\(M_1^j M_2^k+M_1^k M_2^j\)\right]\ln^4\frac{\xi_p}{\yb},
\end{align}
where $c_{i,y}^{\rm PL}$ are the coefficients which appear in the expression
of the impact factor in the pointlike limit Eq.~(\ref{eq:ifpt}).

Equation~(\ref{eq:impactpfour}) thus indeed shows that at LL$x$ level
a collinear log appears to all orders,  but with a fixed power: 
to all orders at LL$x$ the
highest power of log is four. The log  originates from the dynamics
of the quark loop, but it is to all orders
proportional to the pointlike result.  Because
the highest power of the $\ln \pt^2/m_b^2$ terms is fixed
at LL$x$, from our result we cannot exclude  higher order logs and
their 
exponentiation at the
subleading log-$x$ level.

\section{Phenomenology}
\label{sec:hadro}

We now turn to the phenomenological implications of our
results. First, we repeat the comparisons that were presented in the
previous section at the hadronic level. In particular, we validate the
high-energy approximation at leading and next-to-leading order, and
then provide prediction for the transverse momentum distribution at
NLO based on the high-energy approximation.  

As explained in Sect.~\ref{subsec:HEVSLO}, the high-energy
approximation is mostly relevant in the region $\pt>m_H$, where the
pointlike approximation fails, while for lower $\pt$ values the
high-energy result rapidly approaches its pointlike limit, and eventually, for low
enough $\pt$, Sudakov resummation of transverse momentum logs becomes necessary. In the region of interest for this study, as
demonstrated in Sect.~\ref{subsec:comparison}, the contribution of
the bottom quark is entirely negligible. Therefore, in the remainder
of this section we will only include the top contribution.
Furthermore, as previously mentioned, the LL$x$ behaviour of all
partonic channels can be deduced from the gluon-gluon case, and will
thus be included throughout
this section.
All plots are produced with $\mur^2=\muf^2=Q^2$ and
with the PDF4LHC15 NNLO set of parton distributions
{\tt
  PDF4LHC15\_nnlo\_100}~\cite{Butterworth:2015oua,Ball:2014uwa,Harland-Lang:2014zoa,Dulat:2015mca,Carrazza:2015aoa,Gao:2013bia,Watt:2012tq},
for the LHC with $\sqrt{s}=13$~TeV.

\subsection{Validation of the high-energy approximation}
\label{sub:valid}

We have seen in Sect.~\ref{subsec:HEVSLO}  that the pointlike
approximation to the exact result of Ref.~\cite{Baur:1989cm}
deteriorates by an increasingly large amount as $\pt$ grows beyond
$\pt\gsim m_H$, while the high-energy approximation has an accuracy
which is essentially independent of $\pt$ for fixed 
value of the partonic scaling variable $x'$ Eq.~(\ref{eq:xip}). The
partonic $x'$ is of course bounded by the hadronic $\tau'$
Eq.~(\ref{eq:tauprime}), which in turn depends on the scale $Q^2$
Eq.~(\ref{eq:physcale}) which for large $\pt$ is $Q^2\sim
\frac{4\pt^2}{s}$. Because we have seen that the high-energy
approximation is good for $x'\lsim 0.5$ and only deteriorates slowly
for larger values of $x'$, noting that $\tau'=0.5$ corresponds to
$\pt\sim 4.6$~TeV for the LHC at 13 TeV, we expect the high-energy approximation to be
reasonably accurate up to large values of $\pt$.

\begin{figure}
\centering
\includegraphics[width=0.6\textwidth]{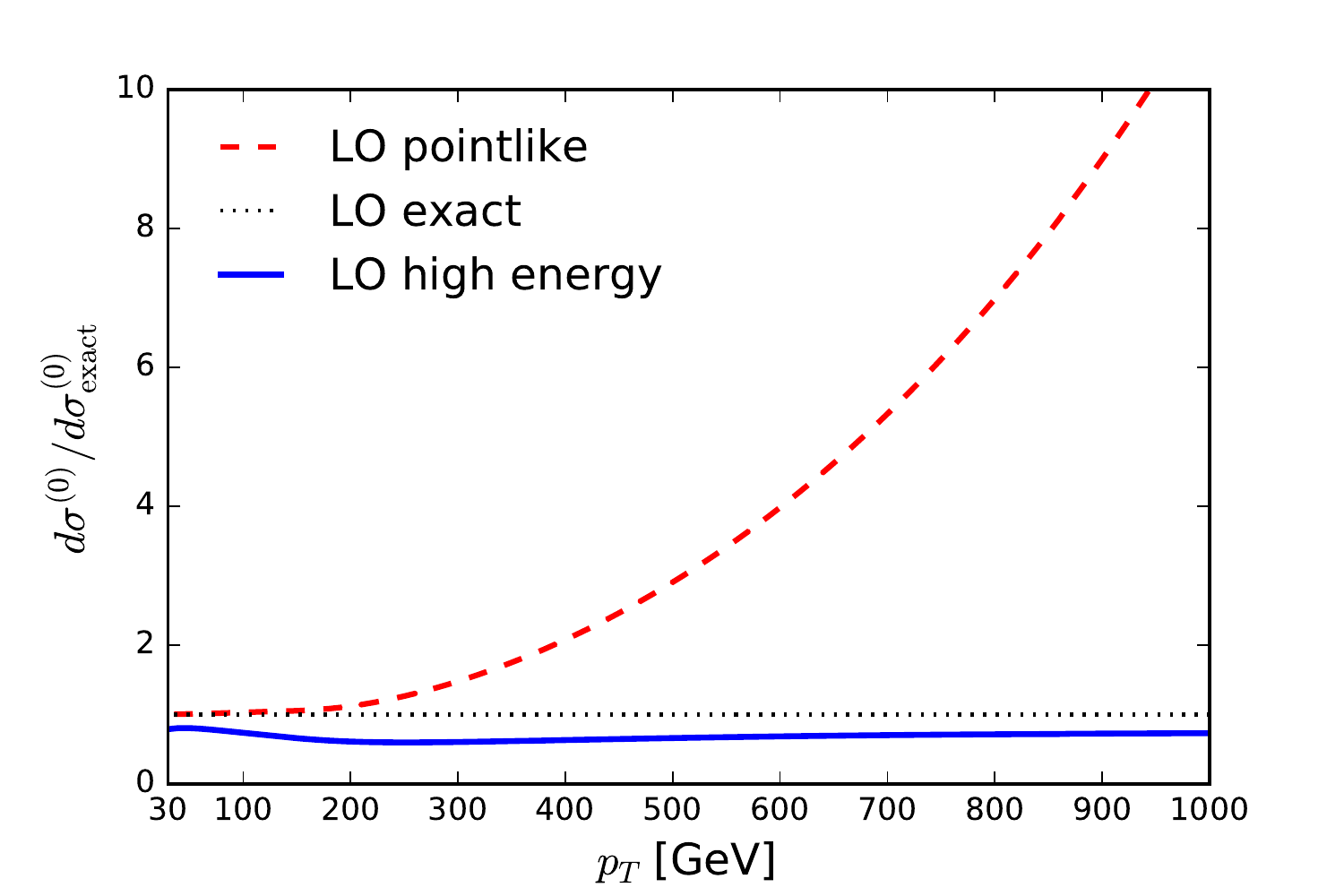}\\
\caption{The ratio of the  high-energy approximation (in solid blue) and of the effective theory result (in dotted red) to the full
  result  for the hadron-level transverse momentum distribution at LO plotted
  as a function of $\pt$ (GeV) at the LHC~13~TeV.}
\label{fig:hevsex}
\end{figure}

We define the NLO transverse momentum distribution
\beq\label{eq:tmom}
\frac{d\sigma}{d\xi_p}\(\tau',\xi_p,\yt,\as\)=\as \frac{d\sigma^{(0)}}{d\xi_p}+\as^2\frac{d\sigma^{(1)}}{d\xi_p}+\Ord\(\as^3\),
\eeq
and the $K$-factor
\beq\label{eq:kdef}
K=1+\frac{d\sigma^{(1)}/d\xi_p}{d\sigma^{(0)}/d\xi_p}.
\eeq
In Figure~\ref{fig:hevsex} we compare the leading order contribution $\frac{d\sigma^{(0)}}{d\xi_p}$ computed in the high-energy approximation  to
the exact result of Ref.~\cite{Baur:1989cm}, and also with the
effective-field theory result. It is clear that, as expected,
the high-energy approximation is most accurate for $\pt\sim m_H$ but
only slowly deteriorates for larger $\pt$: in fact, for all
$0.5\lsim \pt\lsim1$~TeV the high-energy approximation is about 60\% of
the full theory LO result.  The effective field theory result instead is driven
by the fact that at the parton level it has the wrong large-$\pt$
power behaviour, and is off by an increasingly  large factor: at
$\pt\sim1$~TeV it is in fact too large by about one order of
magnitude.

\begin{figure}
\centering
\includegraphics[width=0.49\textwidth]{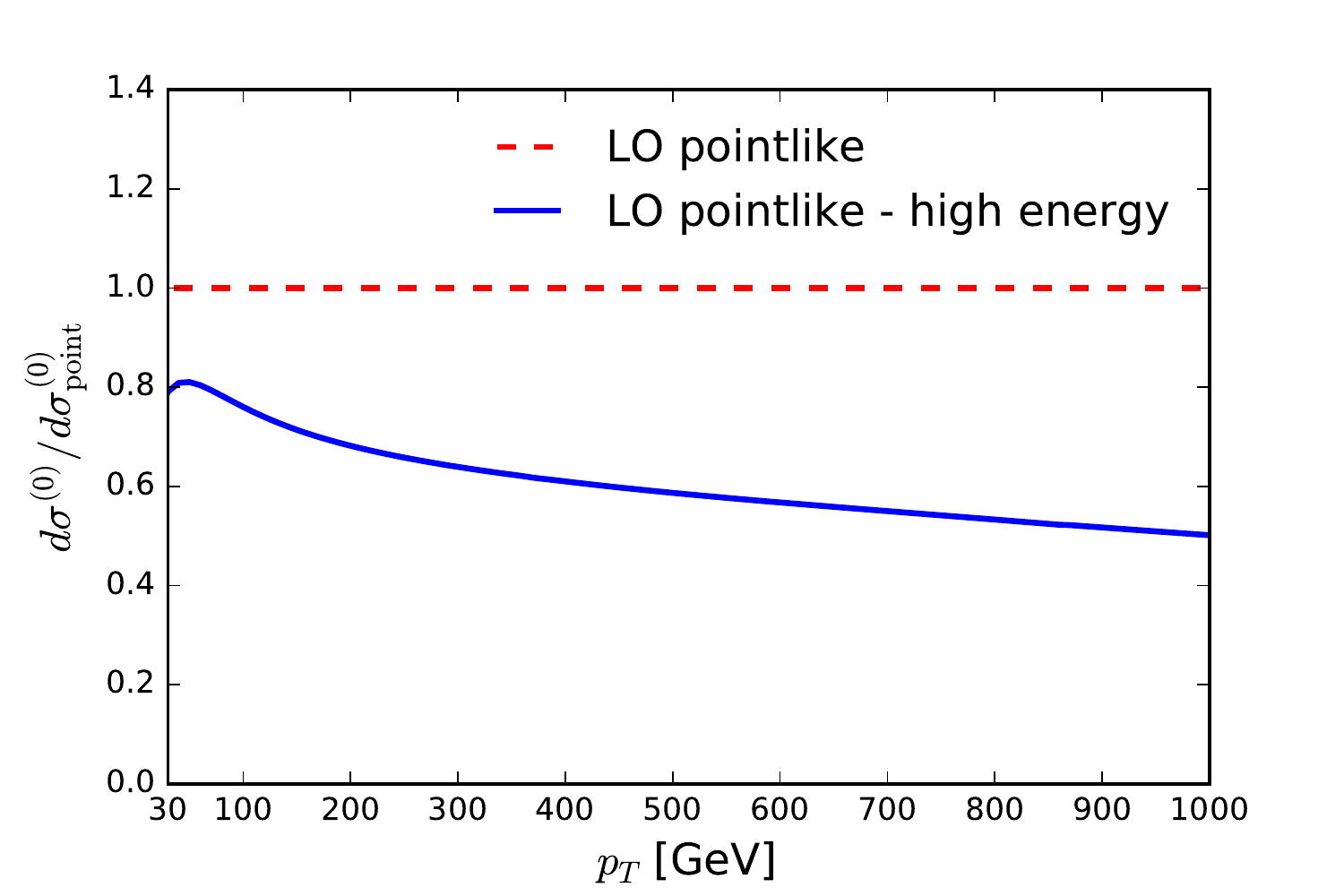}
\includegraphics[width=0.49\textwidth]{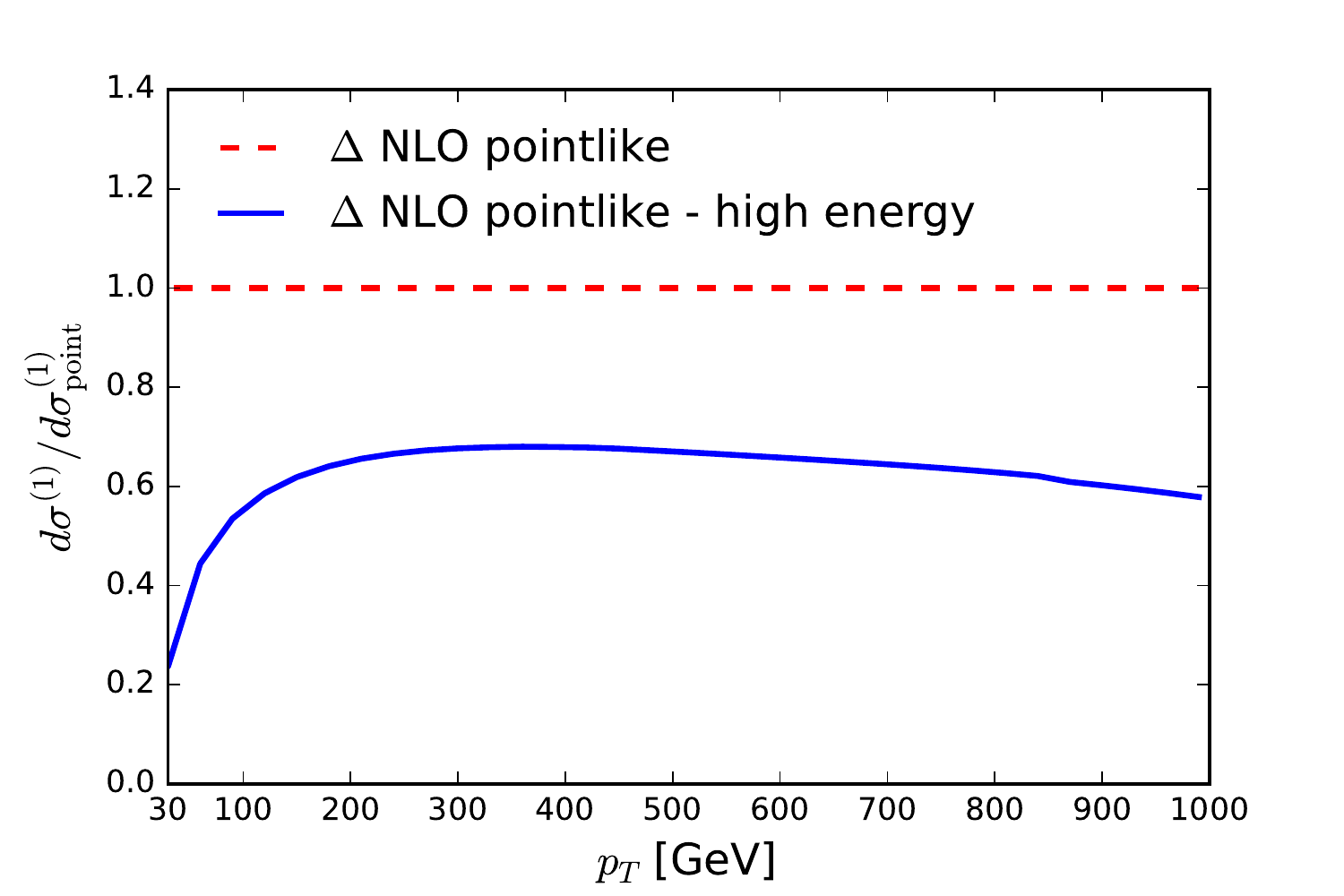}
\caption{The ratio of the  high-energy approximation to the pointlike 
  result for the hadron-level transverse momentum distribution plotted
  as a function of $\pt$ (GeV) at the LHC~13~TeV for the LO, on the left and for
the NLO contribution, on the right.}
\label{fig:hevsex2}
\end{figure}

Beyond leading order we do not have any exact result to compare
to, as only the effective field theory result is available. 
We expect a similar pattern to hold, and we can provide some
evidence for this by studying  the relation between the high-energy
approximation and the full result, both determined in the pointlike
limit. 
This comparison is shown in Fig.~\ref{fig:hevsex2} (left) for the
LO contribution $\frac{d\sigma^{(0)}}{d\xi_p}$. 
It is apparent that the quality of the
high-energy approximation in the pointlike limit is quite similar to
that in the full theory discussed above. 
The NLO contribution $\frac{d\sigma^{(1)}}{d\xi_p}$  is also shown in
Fig.~\ref{fig:hevsex2} (right): we compare
the high-energy
pointlike result of Ref.~\cite{Forte:2015gve} to the full result  of
Ref.~\cite{Glosser:2002gm}. Again, in the medium-high $\pt$ region
we are interested in 
the pattern is quite similar to that seen at LO.

\begin{figure}
\centering
\includegraphics[width=0.7\textwidth]{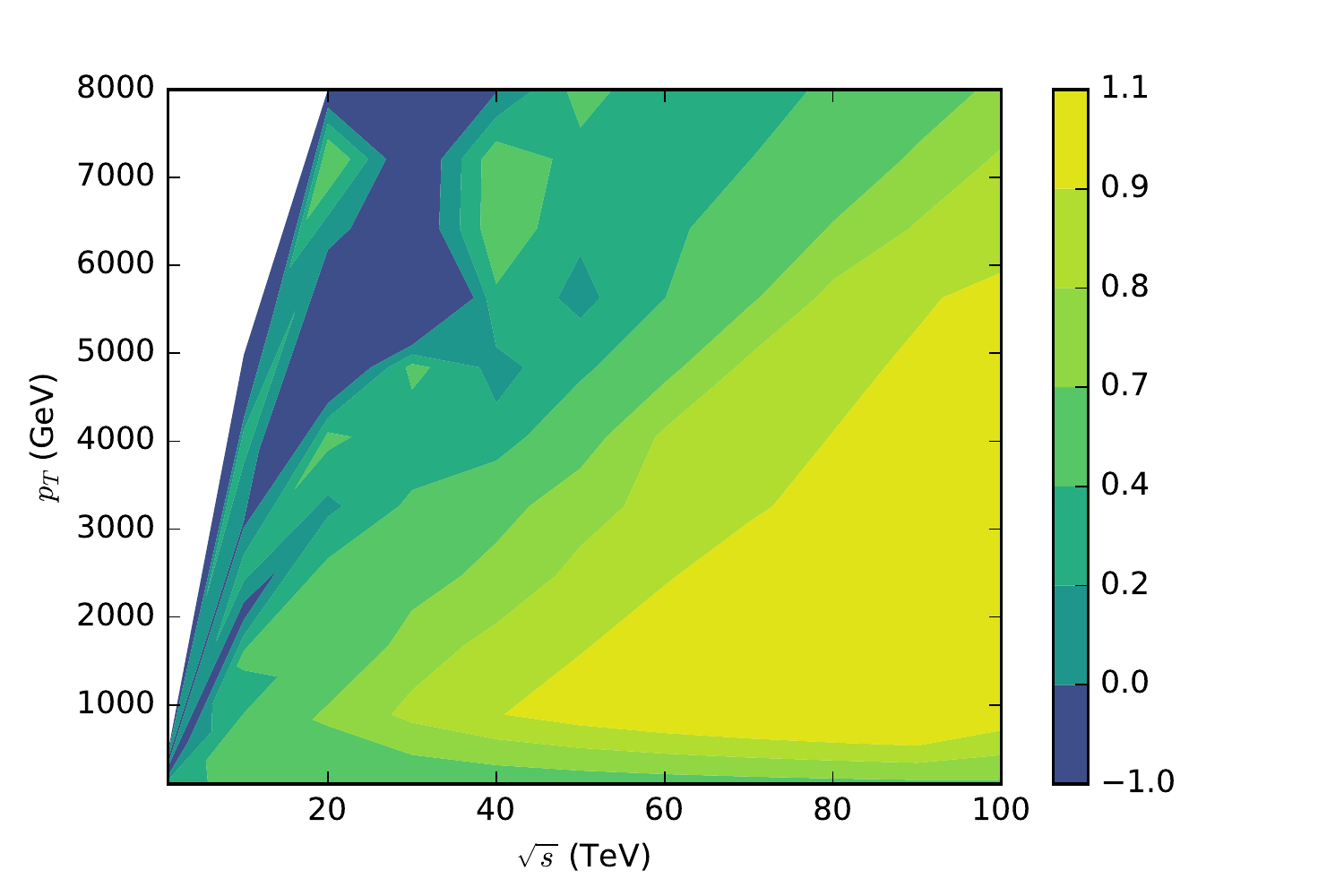}

\caption{The ratio of the NLO contribution
  $\frac{d\sigma^{(1)}}{d\xi_p}$ in the high-energy approximation to exact result, both computed in the pointlike limit, for the Higgs
  transverse momentum distribution at a proton-proton
   collider plotted
  as a function of the transverse momentum $\pt$ (in GeV) and  the
  center-of-mass energy $\sqrt{s}$ (in TeV).}
\label{fig:contourplot}
\end{figure}

This suggests that the high-energy approximation might remain accurate
in a relatively wide kinematic region. In order to test this, we have
repeated the comparison of the high-energy to the full
result for the NLO term $\frac{d\sigma^{(1)}}{d\xi_p}$ , both in the pointlike limit, shown in Fig.~\ref{fig:hevsex2},
for a wide range of values of  $\pt$ and the collider energy. Results are
shown in Fig.~\ref{fig:contourplot}. As expected, the high-energy
approximation becomes better 
as the center-of-mass energy is increased at fixed $\pt$. On the other
hand, if $\pt$ is varied at fixed energy the quality of the
approximation remains constant in a wide range of transverse momenta,
and it only starts deteriorating when the transverse momentum is 
larger than say $\sim20\%$ of its upper kinematic limit
$\sqrt{s}/2$. This is expected because the high-energy limit holds
when $\sqrt{s}$ is much larger than all other scales: for instance,
at large  $\pt$ there are $\ln \pt$ contributions 
which should be resummed to all
orders~\cite{Berger:2001wr}, but are increasingly subleading in the
high-energy expansion. However, in this region the transverse momentum
distribution is tiny, so in practice the high-energy approximation  is
uniformly accurate  throughout the physically relevant region.

\subsection{The mass-dependent spectrum beyond leading order}
\label{subsec:blo}

\begin{figure}
\centering
\includegraphics[width=0.8\textwidth]{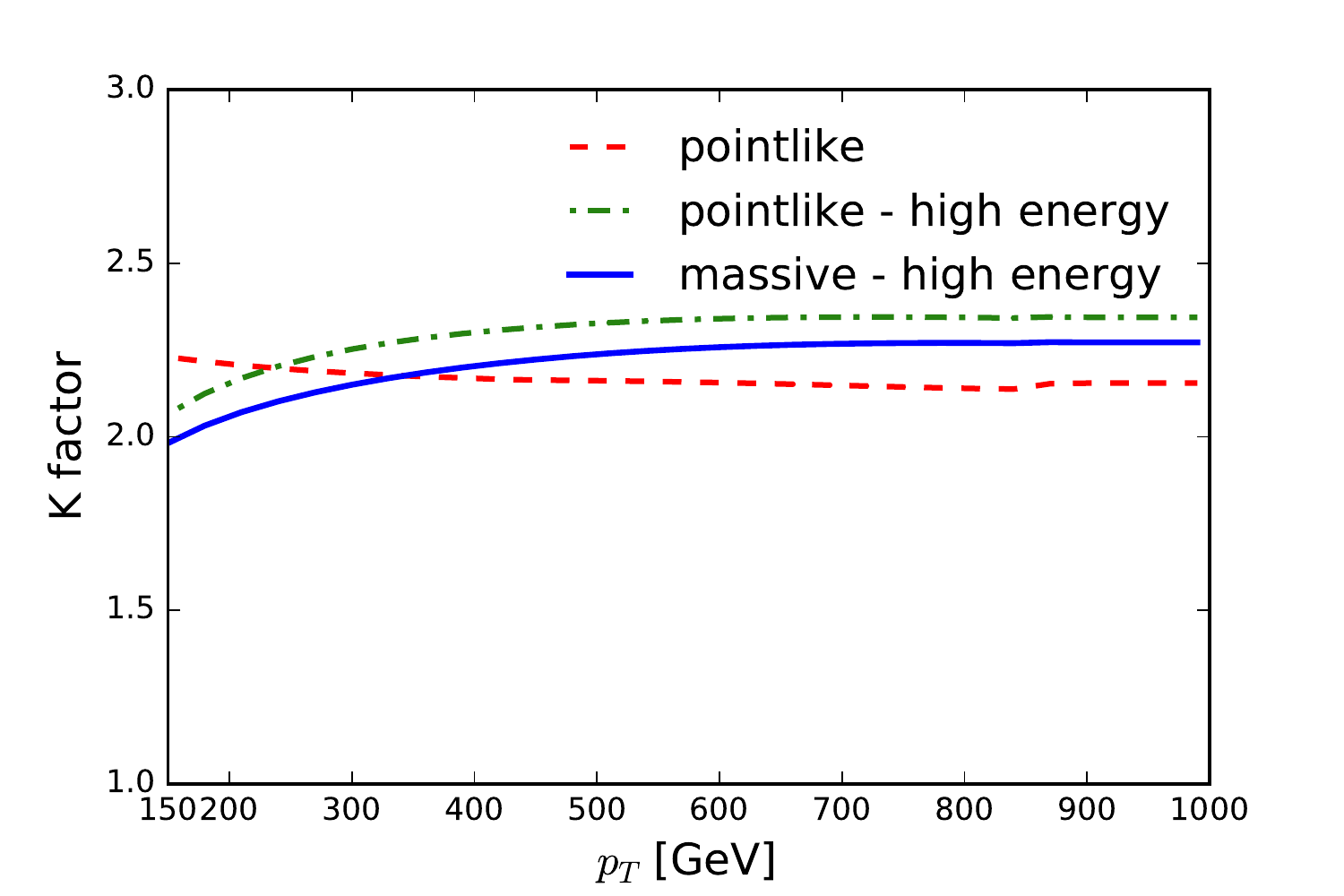}
\caption{The NLO $K$-factor Eq.~(\ref{eq:kdef}) computed using the
  full result in the pointlike limit (red, dashed), and the
  high-energy approximation, either with full mass dependence (blue,
  solid) or in the pointlike limit (green,
  dotdashed). In each case, the LO cross-section is computed
  using the same approximation as the NLO term.}
\label{fig:kkdist}
\end{figure}

We now finally turn to the $\pt$ spectrum of the Higgs boson with
finite top mass beyond leading order. In this case the
exact result is unknown, and thus we can only compare different
approximations.
In Fig.~\ref{fig:kkdist} we compare three different determinations of
the $K$-factor Eq.~(\ref{eq:kdef}) in the high-$\pt$ region we are
interested in: 
using the full pointlike NLO result, the high-energy
approximation to it (i.e. pointlike, and high-energy), and the
high-energy result, but with full mass dependence. In each case, both
the LO and NLO contributions are computed using the same
approximation. This plot shows that for $\pt\gsim200$~GeV 
all these $K$-factors have a
similar behaviour, and differ by comparable amounts. 

This plot suggests two
main conclusions. First, in the only case in which we can compare the
high-energy approximation to the full result, namely the pointlike
limit,  we see that the
high-energy approximation is quite
good (red vs. green curve in Fig.~\ref{fig:kkdist}), 
with an accuracy of about 20\% or better for all $\pt\gsim200$~GeV,
which does not deteriorate as $\pt$ increases. Second, even though (recall
Sect.~\ref{sec:parton}) the
shape of the distribution at high $\pt$ differs between the pointlike and
massive case (a different power of $\pt$) the $K$ factors are similar and
approximately $\pt$ independent, at least in the only case in
which we can compare the pointlike and  massive results, namely the
high-energy limit (green and blue curve).

\begin{figure}
\centering
\includegraphics[width=0.8\textwidth]{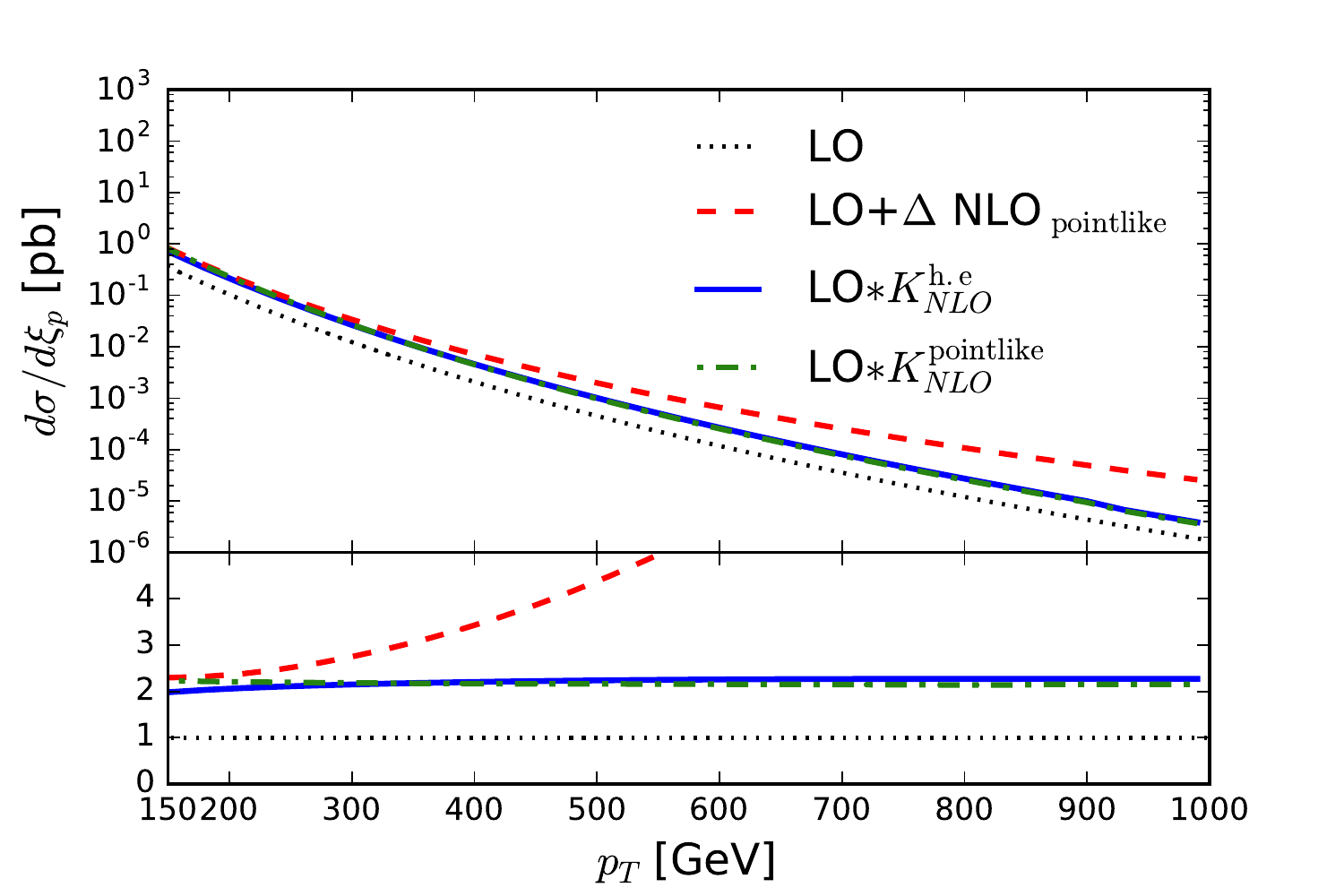}
\caption{Various approximations to the NLO Higgs transverse momentum
  distribution. The curves shown correspond (from top to bottom) 
to adding the pointlike
  approximation to the NLO contribution to the full LO result (red,
  dashed), or to multiplying the full LO result by the $K$-factors of 
Fig.~(\ref{fig:kkdist})  computed respectively in the high-energy
approximation but with full mass dependence (blue, solid) or in the
pointlike approximation (green dot-dashed). The full LO result is also
shown for comparison (black, dotted).
In the bottom plot all curves are shown as ratios  to the exact LO result.}
\label{fig:kdist}
\end{figure}

These two observations, taken together, suggest that the best
approximation to the full NLO result can be obtained by combining the
full LO result with a $K$-factor computed in the high-energy
approximation, namely, by multiplying the LO cross-section by the $K$
factor (blue curve) of Fig.~\ref{fig:kkdist}, corresponding to the high-energy
fully massive result.
 This is our preferred approximation, and it 
is shown in  Fig.~\ref{fig:kdist}, where it is also
 compared to the LO exact result and to the NLO pointlike
 approximation; all results are also shown as ratios to the LO.
It is clear that the pointlike result has the wrong power behaviour
at large $\pt$ and thus fails for $\pt\gsim200$~GeV. 

The comparison of
$K$-factors of Fig.~\ref{fig:kkdist} suggests that if one wishes to
use the NLO pointlike result, rather than the high-energy
approximation,  a better approximation can be obtained by using the
pointlike NLO to compute the $K$ factor (red curve of
Fig.~\ref{fig:kkdist}),  and using this $K$ factor to
rescale the full massive leading order. The quality of this
approximation is possibly comparable to that of our favorite
approximation based on the high-energy limit: indeed, as discussed in
Sect.~\ref{subsec:comparison} this approximation captures the leading
log contributions proportional to $c_0$ in
Eq.~(\ref{eq:coefffulltheory}). This curve is also shown in
Fig.~\ref{fig:kdist}: it is seen to be quite close to our favorite
approximation in a wide range of $\pt$ but it starts departing from it
only at the largest $\pt$ where we expect the high-energy approximation to
be more accurate.

If our approximation to the $K$-factor based on the high-energy limit
is used, it  is natural to ask what is the associated
uncertainty. Having observed that, at the level of $K$-factors, the
difference between the pointlike and massive cases is somewhat smaller
than the difference between the high-energy and full results (see
Fig.~\ref{fig:kdist}), we can conservatively estimate the uncertainty
on the high-energy approximation to be given by the percentage
discrepancy between high-energy and full results (both pointlike)
shown in Fig.~\ref{fig:contourplot}. Of course, this is just the
uncertainty related to the high-energy approximation, which will then
have to be supplemented with all other sources of uncertainty (missing
higher orders, $\alpha_s$, PDFs, etc.).

Before concluding, let us comment on different approaches that can be
found in the literature.  So far studies of finite top mass effects
have been performed by merging different hard-jet multiplicities and
parton showers~\cite{Buschmann:2014sia} and, more recently, in
Ref.~\cite{Frederix:2016cnl} and in the context of 
jet veto analysis~\cite{Banfi:2013eda} and NNLO matching to
parton showers~\cite{Hamilton:2015nsa}.

In Refs.~\cite{Harlander:2012hf,Neumann:2014nha}, finite top
mass effects were evaluated using an asymptotic expansion in inverse
powers of the top mass. This expansion is accurate below $2 m_t$ and
finite-top mass corrections in this region were found to below
10\%. Our approximation, which is valid in the high-$\pt$ region, is
therefore complementary and one would expect that a combination of the
two approaches, in analogy to what was done for the inclusive
case~\cite{Harlander:2009my,Pak:2009dg}, will provide a reliable
approximation across a wide range of $\pt$.  

In Ref.~\cite{Buschmann:2014sia}, top mass effects on the transverse
momentum distribution were calculated using a matched parton shower
approach. This analysis is particularly interesting
for us because both the approach of Ref.~\cite{Buschmann:2014sia}
and ours implicitly relies on the assumption that real radiation provides
the bulk of radiative corrections in the high $\pt$ region.
Nevertheless, this assumption is then used quite differently in the
merged sample and high-energy approximations. Indeed, in the 
former real emission diagrams are accounted for exactly, while
virtual corrections are dropped altogether. The final result
is then affected by merging ambiguities. In the high-energy
approach instead real emission is only included in the LL$x$ approximation,
accompanied however by a matching set of virtual corrections to
ensure a well-defined NLO result. 

Despite these differences, both approaches are supposed to capture the
bulk of NLO corrections in the high $\pt$ region, where the dominance
of real emission is a reasonable assumption.  As a consequence, a
significant disagreement between our results and
Ref.~\cite{Buschmann:2014sia} would imply the presence of large out of
control subleading effects, which would somewhat hamper the
phenomenological relevance of these analysis. Fortunately, it turns
out that the two approaches lead instead to the same conclusions.
Indeed, in the high transverse momentum region we find that the $K$
factor in the pointlike and exact theory are comparable, and that the
shape in $\pt$ of our NLO approximation closely follows the behaviour
of exact LO, in agreement with conclusion drawn with the analysis of
Ref.~\cite{Buschmann:2014sia} (see for example Fig. $4$ of that
reference).

\section{Conclusion}
\label{sec:conclusion}

In this paper, we have applied the high-energy resummation of
transverse momentum distributions of Ref.~\cite{Forte:2015gve} to
Higgs production in gluon fusion with full dependence on heavy quark
masses. We have determined explicit expressions for the resummation
coefficients of the resummed results to all orders. 

The all-order
expression has enabled us to show that the collinear bottom mass logs
which are relevant in the region $m_b<\pt<m_H$ are present to all
orders in the high-energy limit, but with a fixed power of log. We
have then studied the impact of finite mass corrections in the first
few orders. We have shown that the pointlike approximation fails badly
for $\pt \gsim m_t$, while the high-energy approximation provides
reasonably accurate results for center-of-mass energies above a few
TeV and for all $\pt$. Its accuracy does not deteriorate as $\pt$ grows,
unless $\pt$ becomes a sizable fraction of the center-of-mass
energy. 

We have thus argued that the best approximation to the
transverse momentum distribution at past and future LHC energies for
all $\pt \gsim200$~GeV
can be obtained by combining
the known exact leading order result with a $K$-factor computed in the
high-energy approximation. 
At the hadronic level, we have provided results to 
NLO;  the partonic NNLO results presented here suggest that it will be
interesting to investigate the relative accuracy of various
approximations at NNLO and beyond.
More accurate approximations to the full result could be constructed
by combining information on the $\pt$ distribution coming from the
high-energy limit 
with that from the opposite soft limit, in which resummed results are
also available~\cite{deFlorian:2005fzc}. All these developments are
under investigation and will be the object of forthcoming publications.

\section*{Acknowledgements}
We thank R.~Ball, G.~Ferrera, K.~Melnikov, M.~Sch\"onherr, T.Neumann and G.~Zanderighi for
discussions.
S.~F. is supported by the Executive Research Agency (REA) of the European Commission under the Grant Agreement PITN-GA-2012-316704. (HiggsTools). S.~M. is  supported  by the National Science Foundation under Grant No. NSF PHY-0969510, the LHC Theory Initiative, and under Grant No. NSF PHY11-25915.
\appendix

\section{Form factors and perturbative coefficients}
\label{app:impact}
We give here the expressions used in the computation of the
$\pt$-impact factor presented in Sect.~\ref{sec:resummation}.
 We also provide analytic form of the first LO coefficient of the
 expansion in power of $\as$ of the $\pt$-impact factor, discussed in Sect.~\ref{subsec:HEVSLO}.

The $\pt$-impact factor is expressed in Eq.~\eqref{eq:hfinal} as a
double integral over  $\xi_1$ and $\xi_2$ of a function
$F\(\xi_1,\xi_2,\xi_p,\{y_i\}\)$. This function is deduced from the off-shell form factor $\tilde{F}\(\xi,\bar{\xi},\xi_p,\{y_i\}\)$ as
\beq
F\(\xi_1,\xi_2,\xi_p,\{y_i\}\)= \tilde{F}\(\xi_p\,\xi_1,\xi_p\,\xi_2,\xi_p,\{y_i\}\).
\eeq
This form factor is given by~\cite{DelDuca:2001fn,Marzani:2008az}:
\beq\label{eq:ftildeexp}
\tilde{F}\(\xi,\bar{\xi},\xi_p,\{y_i\}\)=\frac{2304 \pi^4}{\abs{\sum_i K\(y_i\)}^2}\abs{\sum_i y_i\,A\(\xi,\bar{\xi},\xi_p,y_i\)}^2
\eeq
with the sum $i$ which runs over the set $\{y_i\}$ of quarks circulating in the loop, and
\begin{align}
&A\(\xi,\bar{\xi},\xi_p, y\)=\frac{C_0\(\xi,\bar{\xi},y\)}{\sqrt{\xi \bar{\xi}}}\notag\\
&\Bigg[\(\frac{2y}{\Delta_3}+\frac{6\xi\bar{\xi}}{\Delta_3^2}\)\(\(\xi_p-\xi-\bar{\xi}\)\(1+\xi+\bar{\xi}\)+4\xi\bar{\xi}\)-\frac{\xi_p-\xi-\bar{\xi}}{2}+2\frac{\xi\bar{\xi}\(1-\xi_p\)}{\Delta_3}\Bigg]\notag\\
&-\frac{1}{\sqrt{\xi\bar{\xi}}}\left[B_0\(-\bar{\xi},y\)-B_0\(1,y\)\right]\left[-\frac{\bar{\xi}}{\Delta_3}\(\xi_p-\bar{\xi}+\xi\)+\frac{6\xi \bar{\xi}}{\Delta_3^2}\(1+\xi_p\)\(1+\xi-\bar{\xi}\)\right]\notag\\
&-\frac{1}{\sqrt{\xi\bar{\xi}}}\left[B_0\(-\xi,y\)-B_0\(1,y\)\right]\left[-\frac{\xi}{\Delta_3}\(\xi_p-\xi+\bar{\xi}\)+\frac{6\xi \bar{\xi}}{\Delta_3^2}\(1+\xi_p\)\(1+\bar{\xi}-\xi\)\right]\notag\\
&+\frac{1}{4\pi^2}\frac{1}{\Delta_3}\frac{1}{\sqrt{\xi \bar{\xi}}}\(\(\xi_p-\xi-\bar{\xi}\)\(1+\xi+\bar{\xi}\)+\xi \bar{\xi}\)
\end{align} 
where $\Delta_3=\(1+\xi+\barxi\)^2-4\xi\barxi$ and
\begin{align}
B_0\(\rho,y\)&=-\frac{1}{16\pi^2}\sqrt{\frac{\rho-4y}{\rho}}\ln\frac{\sqrt{\frac{\rho-4 y}{\rho}}+1}{\sqrt{\frac{\rho-4y}{\rho}}-1}\\
&\notag\\
C_0\(\xi,\barxi,y\)&=\frac{1}{16\pi^2}\frac{1}{\sqrt{\Delta_3}}\Bigg[\ln\(1-y_-\)\ln\(\frac{1-y_-\delta_1^+}{1-y_-\delta_1^-}\)\notag\\
&+\ln\(1-x_-\)\ln\(\frac{1-x_-\delta_2^+}{1-x_-\delta_2^-}\)+\ln\(1-z_-\)\ln\(\frac{1-z_-\delta_3^+}{1-z_-\delta_3^-}\)\notag\\
&+\Li_2\(y_+\delta_1^+\)+\Li_2\(y_-\delta_1^+\)-\Li_2\(y_+\delta_1^-\)-\Li_2\(y_-\delta_1^-\)\notag\\
&+\Li_2\(x_+\delta_2^+\)+\Li_2\(x_-\delta_2^+\)-\Li_2\(x_+\delta_2^-\)-\Li_2\(x_-\delta_2^-\)\notag\\
&+\Li_2\(z_+\delta_3^+\)+\Li_2\(z_-\delta_3^+\)-\Li_2\(z_+\delta_3^-\)-\Li_2\(z_-\delta_3^-\)\Bigg]
\end{align}
with
\begin{align}
\delta_1&\equiv\frac{-\xi+\barxi-1}{\sqrt{\Delta_3}}, & \delta_2&\equiv \frac{\xi-\barxi-1}{\sqrt{\Delta_3}}, & \delta_3&\equiv\frac{\xi+\barxi+1}{\sqrt{\Delta_3}},\\
&&\delta_i^\pm&\equiv\frac{1\pm\delta_i}{2}, &&\\
\end{align}
and
\begin{align}
x_\pm&\equiv -\frac{\barxi}{2y}\Bigg(1\pm\sqrt{1+\frac{4y}{\barxi}}\Bigg),\\
y_\pm&\equiv -\frac{\xi}{2y}\Bigg(1\pm\sqrt{1+\frac{4y}{\xi}}\Bigg),\\
z_\pm&\equiv \frac{1}{2y}\Bigg(1\pm\sqrt{1-4y}\Bigg).
\end{align}
The form factor $A$ can be expressed in terms of standard
  one-loop scalar integrals~\cite{Ellis:2007qk} by letting
  $C_0(\xi,\bar\xi,y_i) =m_h^2 I_3(-\xi m_h^2,-\bar\xi  m_h^2,m_h^2,m_i^2,m_i^2,m_i^2)/(16\pi^2)$ 
  and $B_0(\rho,y)-B_0(1,y)=\left[I_2(\rho m_h^2,m_i^2,m_i^2)-I_2( m_h^2,m_i^2,m_i^2)\right]/(16\pi^2)$.
As already stated in the main text, the analytic continuation of the form
factor has to be handled by giving $y$ a small negative imaginary part.

Using these expressions, we obtain the following limiting cases
\begin{align}
\label{eq:yinfty}
\lim_{y\to\infty} F\(\xi_1,\xi_2,\xi_p,y\)&=\frac{\(1-\xi_1-\xi_2\)^2}{4\xi_1 \xi_2}\\
\label{eq:xip0}
\lim_{\xi_p\to 0} F\(\xi_1,\xi_2,\xi_p,\{y_i\}\)&=\frac{\(1-\xi_1-\xi_2\)^2}{4\xi_1 \xi_2}.
\end{align}

Finally, we provide an analytic expression for the first expansion
coefficient $c_0$ Eq.~(\ref{eq:cjk}) of the perturbative expansion
Eq.~\eqref{eq:hexpansion2}: 
\beq
\label{eq:F10}
c_0\(\xi_p,\{y_i\}\)=\frac{2304 \pi^4}{\abs{\sum_i K\(y_i\)}^2}\abs{\sum_i y_i\,A\(0,\xi_p,\xi_p,y_i\)}^2
\eeq
with
\begin{align}
\label{eq:A1xp0}
A\(0,\xi_p,\xi_p,y\)&=\frac{1}{32\pi^2}\Bigg(\frac{4y-1-\xi_p}{\(1+\xi_p\)^2}\Bigg[\ln^2\frac{\sqrt{1-4y}-1}{\sqrt{1-4y}+1}-\ln^2\frac{\sqrt{1+\frac{4y}{\xi_p}}-1}{\sqrt{1+\frac{4y}{\xi_p}}+1}\Bigg]\notag\\
&+\frac{4\xi_p}{\(1+\xi_p\)^2}\Bigg[\sqrt{1-4y}\ln\frac{\sqrt{1-4y}+1}{\sqrt{1-4y}-1}-\sqrt{1+\frac{4y}{\xi_p}}\ln\frac{\sqrt{1+\frac{4y}{\xi_p}}+1}{\sqrt{1+\frac{4y}{\xi_p}}-1}\Bigg]\notag\\
&+\frac{4}{1+\xi_p}\Bigg).
\end{align}

%%%%%%%%%%%%%%%%%%%%%
\phantomsection
\addcontentsline{toc}{section}{References}
%
%\bibliographystyle{jhep}
%\bibliography{biblio}
\input{HiggsMassQuarkf.bbl}
\end{document}

%% file: HiggsMassQuarkf.bbl
\providecommand{\href}[2]{#2}\begingroup\raggedright\endgroup